\documentclass[aps,twocolumn,superscriptaddress,nofootinbib,floatfix]{revtex4-2}

\usepackage{amsmath}
\usepackage{mathrsfs}
\usepackage{amssymb}
\usepackage{bm} 
\usepackage{graphicx,grffile,textcomp}
\usepackage{dcolumn} 
\usepackage[usenames,dvipsnames]{xcolor}
\usepackage{multirow}
\usepackage{epstopdf}
\usepackage{mathtools} 
\usepackage{tabularx}
\usepackage{microtype}
\usepackage{comment}
\usepackage[version=4]{mhchem} 
\usepackage{array}
\usepackage{float}
\usepackage{wrapfig} 
\usepackage{url}
\usepackage{hyperref}
\usepackage{cleveref}
\usepackage[utf8]{inputenc}
\usepackage[T1]{fontenc}
\usepackage{bbold}

\hypersetup{colorlinks=true, linkcolor=blue, urlcolor=blue, citecolor=blue}

\definecolor{myred}{RGB}{230, 7, 77}

\definecolor{blue(ryb)}{rgb}{0.01, 0.28, 1.0}
\definecolor{jade}{rgb}{0.0, 0.66, 0.42}
\definecolor{cadmiumred}{rgb}{0.89, 0.0, 0.13}
\definecolor{darkviolet}{rgb}{0.58, 0.0, 0.83}

\begin{document}

\title{Curvature effects in interfacial acidity of amphiphilic vesicles}

\author{Petch Khunpetch}
\email{petch.k@rumail.ru.ac.th}
\thanks{Co-first and co-corresponding author}
\affiliation{Department of Physics, Faculty of Science, Ramkhamhaeng University, Bang Kapi, Bangkok, Thailand \&
School of Physical Sciences, University of Chinese Academy of Sciences, Beijing, China}

\author{Arghya Majee}
\email{majee@pks.mpg.de}
\thanks{Co-first and co-corresponding author}
\affiliation{Max Planck Institute for the Physics of Complex Systems, 01187 Dresden, Germany}

\author{Hu Ruixuan}
\affiliation{School of Physical Sciences, University of Chinese Academy of Sciences, Beijing, China}

\author{Rudolf Podgornik}
\affiliation{School of Physical Sciences, University of Chinese Academy of Sciences, Beijing, China \&
Kavli Institute for Theoretical Sciences, University of Chinese Academy of Sciences, Beijing, China \& CAS Key Laboratory of Soft Matter Physics,  Institute of Physics, Chinese Academy of Sciences,  Beijing, China \& Wenzhou Institute of the University of   Chinese Academy of Sciences,  Wenzhou, Zhejiang, China}

\date{May 18, 2023}

\begin{abstract}
We analyze the changes in the vicinal acidity 
(pH) at a spherical amphiphilic 
membrane. The membrane is assumed to 
contain solvent accessible, embedded, 
dissociable, charge regulated moieties. 
Basing our approach on the linear 
Debye-H\" uckel  as well as the non-linear  
Poisson-Boltzmann theory, together with 
the general Frumkin-Fowler-Guggenheim 
adsorption isotherm model of the charge 
regulation process, we analyse and review 
the dependence of the local pH 
on the position, as well as bulk electrolyte 
concentration, bulk pH and curvature 
of the amphiphilic single membrane vesicle. 
With appropriately chosen adsorption parameters 
of the charge regulation model, we find a good 
agreement with available experimental data.
\end{abstract}

\maketitle

\section{Introduction\label{Sec:1}}

The charging state of phospholipid membranes \cite{Cev18}, lipid nanoparticles \cite{Zhdanov2023}, but also other amphiphilic \cite{Jonsson2002} as well as proteinaceous self-assemblies \cite{Zhou2018}, is governed by the protonation/deprotonation equilibria of dissociable surface molecular groups in  contact with the aqueous subphase. In the case of proteins \cite{Simonson_2003} the negative charges stem from the deprotonated carboxylate on the side chains of aspartic and glutamic acid, and the deprotonated hydroxyl of the phenyl group of tyrosine, while the positive charge originates from the protonated amine group of arginine and lysine, as well as the protonated secondary amine of histidine  \cite{Nap14}. In the case of phospholipids  the negative charge is derived from deprotonated phosphate groups and deprotonated carboxylate, while the positive charge, though rare in naturally occuring lipids \cite{Gal21}, stems from protonated amine group or other titratable molecular moieties with an engineered dissociation constant \cite{Khunpetch2022}.  

Among the phenomena in biomolecular assemblies where charging equilibria are particularly important, one can specifically name the electrostatic interactions between membranes \cite{Cev90}, ion transport across the membranes \cite{Pana2012}, as well as the insertion and transolocation of membrane proteins \cite{Sarkar2018}. To these well known examples one could also add the emerging role of charging equilibria in viral proteinaceous capsid shells  \cite{ZANDI20201}, their interactions with various substrates \cite{D1SM00232E} and structural reconstructions and  maturation processes in chimeric protein-lipid capsid shells \cite{Roshal2019, C9SM01335K, D2NA00461E}. 

The charging equilibria in biomacromolecular assemblies typically involve local $\textrm{pH}$ and local bathing solution ion concentrations, which - as has been recognized for a while - in general differ from the bulk conditions \cite{Longo2011,Longo2013,Nap14}, implying that the changes in the bathing solution properties will affect not only the $\textrm{pH}$ sensing and $\textrm{pH}$ response of lipid membranes \cite{Ang18} but will also - and even more importantly - affect the membrane protein(s) entering different biochemical reactions required for the sustainability and proliferation of life. Elucidating the quantitative details of the relation between {\sl bulk} and {\sl local} solution properties thus constitutes one of the challenges in the description of biomacromolecular assemblies.

\begin{figure*}[t]
\centering
\includegraphics[width=7cm]{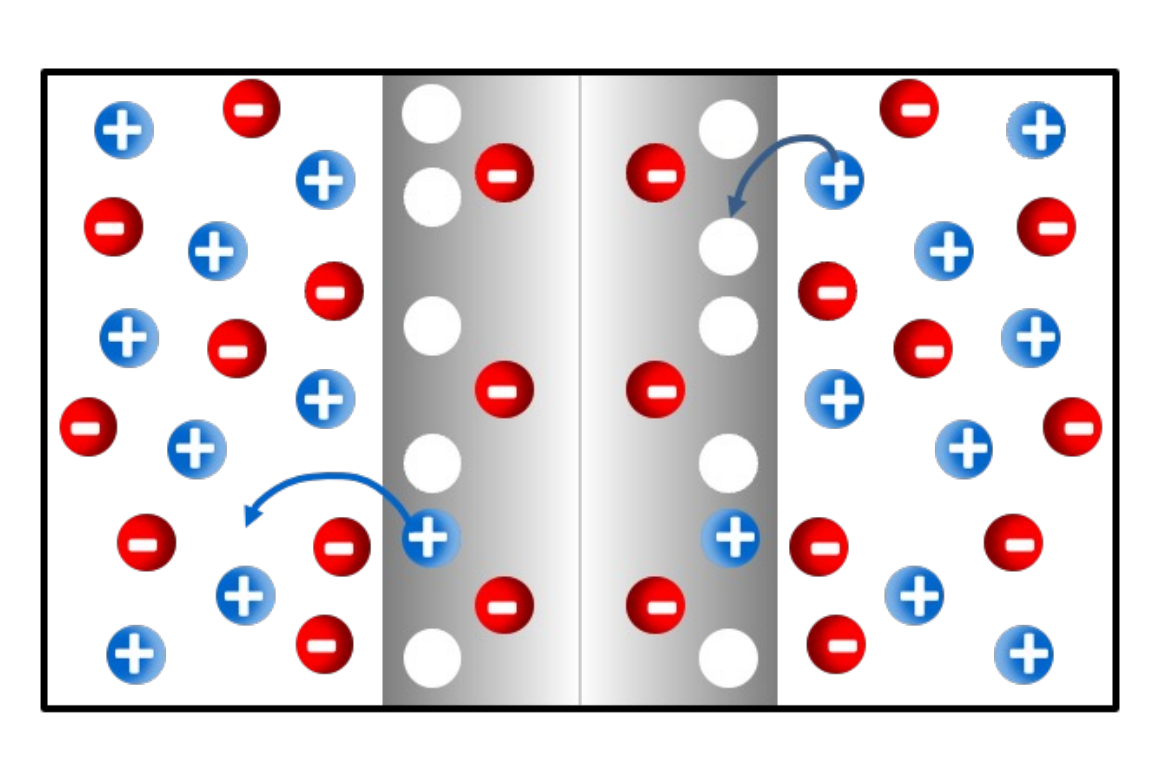}~~~~~~~~~~~~~
\includegraphics[width=3.5cm]{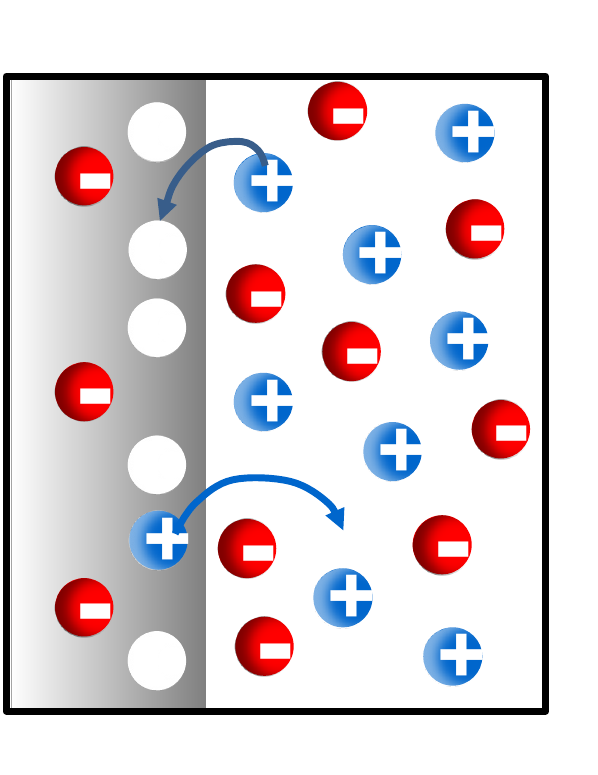} \\
~~~~~~~~~~~~~~~~~~~~~~~~~~~~~~~~~~(a)~~~~~~~~~~~~~~~~~~~~~~~~~~~~~~~~~~~~~~~~~~~~~~~~~~~(b)~~~~~~~~~~~~~~~~~~ \\
~\\
\includegraphics[width=10cm]{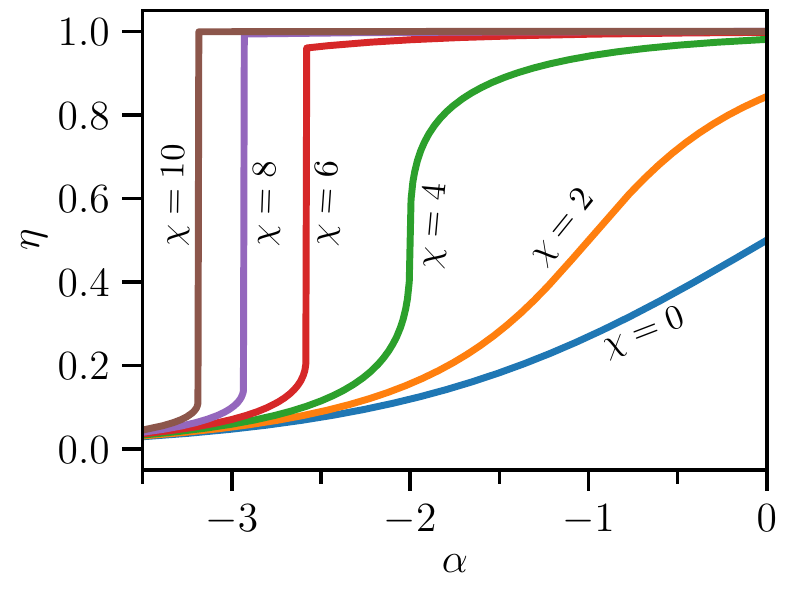}\\
(c)
    \caption{(a) A schematic representation of charge regulation at the surface of an amphiphilic bilayer  containing dissociable protonation/deprotonation sites.
    In model (1), Eq. \eqref{sigma}, the charge is regulated in the interval $- \sigma_0 \leq \sigma_{1,2} \leq \sigma_0$. In the  case of the model (2) Eq.~\eqref{sigma2}
    the charge is regulated in the interval $0 \leq \sigma_{1,2} \leq \sigma_0$. In both cases $\sigma_0$ is a fixed maximal structural  charge density.  Protons provided by the dissociable surface moieties can exchange with the solution and in the process  charge/discharge the dissociable groups. (a) Charge regulated bilayer membrane with CR sites on the surfaces of both monolayers. (b) A single monolayer of the charge  regulating bilayer. (c) Solution of the Frumkin-Fowler-Guggenheim isotherm, Eq.~\eqref{BC2a}, for a single charge regulated monolayer with  different values of the $\chi$ interaction parameter with $\psi = 0$. The critical isotherm is given by $\chi=-2\alpha$, corresponding to $(\alpha, \chi) = (-2,4)$, and below this value there is a discontinuous, first order transition from one charged state to another one. $\chi =0$ corresponds to the Langmuir isotherm with no discontinuous charging state transition. \label{Fig0}}
\end{figure*}

We have recently formulated \cite{Maj18,Maj19,Maj20} a theoretical model that couples the full macroscopic continuum description of electrostatic interactions with the surface protonation/deprotonation reactions of charged lipids and/or other amphiphillic molecules \cite{Khunpetch2022}. This model not only yields the details of the lipid charging state as a function of the curvature and bathing solution parameters, such as $\textrm{pH}$ and salt concentration but also, and this will be the focus here, the full spatial profile of the $\textrm{pH}$ in the vicinity of the membrane. In this way we can connect the interfacial curvature with the  interfacial $\textrm{pH}$ for the non-planar self-assemblies and assess the role of the surface curvature in the interfacial acidity/basicity properties. This connection in itself attests to the fact that the knowledge of the {\sl bulk} bathing solution properties does not imply that we know what the {\sl local} solution properties are, that in the last instance determine the solution state near the proteins and lipids functional groups. Below we will argue that the local $\textrm{pH}$ can actually veer off quite far from the nominal values set in the bulk.  

In what follows we will first recapitulate the basic features of our theoretical model with the Poisson-Boltzmann volume free energy functional for mobile charges and the Frumkin-Fowler-Guggenheim adsorption isotherm model, formulated in terms of the appropriate surface free energy, for the surface charging equilibrium. We then solve the model in the linearized Debye-H\" uckel (DH) approximation as well as in  the full Poisson-Boltzmann (PB) theory for a spherical vesicle with finite thickness permeable membrane, whose solvent accessible surfaces contains the dissociable moieties. We specifically describe the spatial profiles of the vicinal as well as luminal $\textrm{pH}$ as a function of the parameters of the model. We finally comment on the salient features of the interfacial acidity and its dependence on the bulk properties.

\section{Charge Regulation Model\label{Sec:2}}

We consider a spherical vesicle with salt solution on the two sides of its bilayer membrane, composed of two charge-regulated monolayers containing dissociable moieties, as shown in Fig.~\ref{Fig0}. The inner radius of the vesicle shell is $R$ with charge density $\sigma_{1}$ and the outer radius is 
$R+w$ with charge density $\sigma_{2}$. The approach described below broadly follows our previous work and we shall only list the relevant details that were further elaborated in \cite{Maj18,Maj19,Maj20,Khunpetch2022}. We will focus on two CR models corresponding either to a symmetric charge regulation, with the surface charge density $-\sigma_0 \leq \sigma_{1,2} \leq \sigma_0$, or an asymmetric CR models with the surface charge density $0 \leq \sigma_{1,2} \leq \sigma_0$, where $\sigma_0 = e_0 n_0$ is a structural charge parameter corresponding to the maximal dissociated surface charge, with $e_0$ the elementary charge ($e_0 > 0$) and $n_0$ a structural density of dissociable moieties. 

Furthermore, we assume that $\eta_{1,2} \in [0, 1]$ are the fractions of the neutral lipid heads where the
adsorption/desorption (association/dissociation) of as yet unspecified cations can take place on the inner and outer 
monolayers and that $\sigma_{1,2}$ and $\eta_{1,2}$ are  uniform over the two monolayers.  

In model (1) \cite{Maj18} the CR surface charge density is then given by
\begin{equation}
\sigma_i  =  
2 n_i e_0 \left(\eta_i - \frac{1}{2}\right),
\label{sigma}
\end{equation}
with $- e_0 n_{1,2} \leq \sigma_{1,2} \leq e_0 n_{1,2}$ that can obviously change sign, while in model (2) \cite{Nin71} we  assume that the CR surface charge density is given by 
\begin{equation}
\sigma_i  =  n_i e_0 \left(\eta_i - 1 \right),
\label{sigma2}
\end{equation}
so that $-e_0 n_{1,2} \leq \sigma_{1,2} \leq 0$ and consequently cannot change sign. We note that the two models described by Eq.~\eqref{sigma} and Eq.~\eqref{sigma2}  correspond to two situations, where the protonation/deprotonation of dissociable moieties can lead to positive or negative net charge, or it leads only to a single type of net charge. In the first case the model is charge-symmetric while in the second case it is charge-asymmetric. In reality the situation is probably somewhere in between with an additional compositional asymmetry associated with the outer and inner monolayer.

In numerical calculations the static dielectric constant of water is assumed as  $\epsilon_{w}=80$ and 
that of the lipid bilayer membrane as $\epsilon_{p}=5$. The Debye 
screening length $(\lambda_D = \kappa_D^{-1})$ varies from about $0.34\,\textrm{nm}$ to 
$10.75\,\textrm{nm}$, corresponding to the monovalent salt concentration 
ranging from $1-0.001 \textrm{M}$ \cite{Sho16}. Moreover, we define 
$\mu=\epsilon_{p}/\epsilon_{w}$.\textbf{}

\section{Electrostatic free energy: Poisson-Boltzmann and Debye-H\" uckel forms}\label{sec:es}

We start with the standard PB free energy, or the DH free energy in the linearized case, that depends on the distribution of mobile charges, assumed to belong to a univalent electrolyte with a fixed bulk chemical potential. The surface charges are assumed to be located at both interfaces of a spherical membrane.   
Most of our results pertain to the DH approximation that has proven to be useful not only to provide qualitative but also quantitative results in the context of various problems involving interactions of charged colloidal particles \cite{Maj16,Maj18,Maj19,Beb20,Roccia2021,Roccia2022}. We should, however, clearly state that the linearization implied by the DH approximation pertains only to electrostatics but not to the surface charging equilibrium which is always considered in its full, non-linear form.

There are various ways to write down the PB free energy \cite{Mar21} and we choose the field description, with the radially varying  mean-field electrostatic potential $\psi(r)$ as the only relevant variable. The total PB electrostatic free energy is then given by  
\begin{align}
{\mathcal F}_{ES} =& -\!\int\limits_V \! dV \! \left[\frac{\epsilon_w\epsilon_0}{2}\!\left(\frac{d\psi(r)}{dr}\right)^{\!2} \! + 2n_I \left(\cosh{\beta e_0 \psi(r)}-\!1\right)\right] \nonumber\\
&+\oint\limits_{A_1} dA_1~\psi\left(R_1\right) \sigma_1 + \oint\limits_{A_2} dA_2~\psi\left(R_2\right) \sigma_2,
\label{electrostatic1}
\end{align}
where $n_I$ is the univalent electrolyte concentration in the bulk, $R_1 = R$ 
and $R_2 = R + w$, while the equilibrium 
value of $\psi(r)$ is obtained from the corresponding Euler-Lagrange (EL) equation. The volume integral extends over all the regions except the bilayer interior. Within the bilayer interior there are no mobile charges and the second term in the square brackets is absent so that the electrostatic free energy is simply
\begin{align}
\mathcal{F}_{ES} = & - \frac{\epsilon_p\epsilon_0}{2} \int\limits_V dV  \left(\frac{d\psi(r)}{dr}\right)^{\!2} \nonumber\\
&+\oint\limits_{A_1} dA_1~\psi\left(R_1\right) \sigma_1 + \oint\limits_{A_2} dA_2~\psi\left(R_2\right) \sigma_2,
\label{electrostatic11}
\end{align}
where the volume integral now extends over the bilayer interior and of course, the permittivity $\epsilon_p$ needs to be used. 

A common approach to electrostatic effects in soft matter and specifically in the case of charged membrane vesicles is {\sl via} the DH approximation \cite{Muthu2023} often coupled 
together with small curvature, second order expansion \cite{And95, Fog99, Gal21}. In the 
DH approximation, valid strictly for $\beta e_0 \psi(r) \ll 1$ but yielding qualitatively similar results to the full PB solution also outside this limit \cite{Khunpetch2022}, the corresponding expressions 
for the electrostatic free energy Eq.~\eqref{electrostatic1} simplifies considerably to 
\begin{eqnarray}
{\mathcal F}_{ES} & = & -\frac{\epsilon_w\epsilon_0}{2} \int\limits_V dV \Bigg( \left(\frac{d\psi(r)}{dr}\right)^{\!2} + \kappa_D^2  \psi(r)^2\Bigg)\nonumber\\
&& +\oint\limits_{A_1} dA_1~\psi\left(R_1\right) \sigma_1 + \oint\limits_{A_2} dA_2~\psi\left(R_2\right) \sigma_2,
\label{electrostatic2}
\end{eqnarray}
where the inverse square of the Debye screening length $\lambda_D$ is given by 
$\kappa_D^{2} = 2 n_I \beta e_0^2/\left(\epsilon_w\epsilon_0\right)$ and the volume integral
again extends over all the regions except the bilayer interior. While the free energies 
Eq.~\eqref{electrostatic1} and  Eq.~\eqref{electrostatic2} imply the PB and the DH 
equation in the regions accessible to electrolyte ions \cite{Maj16}, respectively, 
Eq.~\eqref{electrostatic11} leads to the standard Laplace equation inside the lipid
dielectric core. Inserting the solution of the EL equations back into  Eq.~\eqref{electrostatic2},
it  is then further reduced to a form corresponding to the Casimir charging process \cite{verwey48a}
\begin{align}
{\mathcal F}_{ES}(\sigma_1, \sigma_2, R)&= 4\pi \sum_{i=1}^2 R_i^2 \int\limits_0^{\sigma_i} \!\!\!d\sigma_i ~\psi(\sigma_1, \sigma_2, R_i)\notag\\
&\longrightarrow \frac{1}{2} \sum\limits_{i=1}^2
4\pi R_i^2 ~ 
\sigma_i~\psi\left(\sigma_1, \sigma_2, R_i\right) ,
\label{nkcajwgscfkj3}
\end{align}
where the right arrow indicates the DH limit of the same expression where the potentials are
linear functions of the charge density. With the explicit solution for the electrostatic
potential, see Appendix  \ref{DHapp},  we can derive the DH expression for the electrostatic
free energy per area as a function of the radius of curvature $R$ to inverse quadratic order
\cite{Sho16}, obtaining an approximate but highly accurate form of the free energy
\begin{align}
\frac{\kappa_D \epsilon_0\epsilon_w \mathcal{F}_{el}(\sigma_1, \sigma_2, R)}{2\pi R^2} =& f_0\left(\sigma_1, \sigma_2, \kappa_D, w \right)\notag\\
&+\frac{f_1\left(\sigma_1, \sigma_2, \kappa_D, w \right)}{\kappa_D R}\notag\\
&+\frac{f_2\left(\sigma_1, \sigma_2, \kappa_D, w \right)}{(\kappa_D R)^2},
\label{fel}
\end{align}
where the curvature independent terms, $f_0\left(\sigma_1, \sigma_2, \kappa_D, w \right),
f_1\left(\sigma_1, \sigma_2, \kappa_D, w \right)$ and $f_2\left(\sigma_1, \sigma_2, \kappa_D, w \right)$
are explicitly given in the Appendix \ref{curvapp}. In general the above free energy density of a curved
membrane is not symmetric in the two solvent accessible surface charge densities that were assumed to be constant.

The DH electrostatic free energy for fixed surface charges displays a general quadratic dependence on the 
curvature of the lipid bilayer. This quadratic dependence of electrostatic free energy was standardly taken
as a point of departure for the electrostatic renormalization of the mechanical properties of membranes,
such as surface tension and bending rigidity \cite{Win88, Mit89, Lek90, Dup90, Har92, Sho16} but ceases
to be the case of charge regulated membranes.

\section{Charge regulation free energy and self-consistent boundary conditions}\label{sec:cr}

Assuming that the inner and outer membrane surfaces are chemically identical we 
presume that the surface charge regulation process can be described by the 
Frumkin-Fowler-Guggenheim adsorption isotherm, which is a two parameter adsorption model \cite{Koo20},
parameterized with  the adsorption energy,  $\alpha$,  the interaction energy between adsorbed ions, $\chi$,
and the lattice gas entropy. For $\chi = 0$ the Frumkin-Fowler-Guggenheim model reduces to the Langmuir model.
Other, multiparametric models of variable complexity can be defined but will not be analyzed here \cite{Bor01}.

The corresponding charge  regulation free energy densities of the inner and outer membrane surfaces denoted by 
$i=1,2$ are given by
\begin{align}
\frac{{\mathcal F}_{CR}(\eta_{i})}{4\pi R_{i}^{2}} = &n_{0} ~k_BT\Big[ - \alpha\eta_{i} -\frac{\chi}{2}\eta_{i}^2\notag\\
&+ \eta_{i}\ln\eta_{i}  + (1-\eta_{i})\ln{(1-\eta_{i})} \Big].
  \label{cr1}
\end{align}
This can be furthermore normalized w.r.t. the inner area $4\pi R^{2}$ which is used later. The first two terms
in the free energy are enthalpic in origin. The other terms are the lattice gas mixing  entropy of charged sites
with the surface area fraction $\eta$ and neutralized sites with the surface area fraction $1-\eta$.

In the case of phospholipids such as Phosphatidic acid (PA, smallest and simplest phospholipid, precursor for
other phospholipids), Phosphatidylserine (PS) and Phosphatidylglycerol (DPPG), the negative charge comes from
deprotonated conjugate base of phosphoric acid and deprotonated carboxylate, while the positive charge comes
from the protonated amine or ammonium head of cationic lipids but can be also substituted with an engineered
dissociation constant.

In these cases of charge regulation the adsorbing/desorbing particles are identified as protons and $\alpha$
is then the deprotonation free energy difference \cite{Avn18} which in the case of the Langmuir adsorption
model \cite{Nin71} becomes
\begin{equation}
\alpha = (\mathrm{p}K_{\mathrm{a}}-\textrm{pH}) \ln 10, 
\end{equation}
where $\textrm{p}K_{\textrm{a}}$ is the dissociation constant of the deprotonation reaction and
$\textrm{pH} = -\log_{10}{[\textrm{H}^{+}]}$ is the acidity. The Langmuir model in this context
is equivalent to a Henderson-Hasselbalch equation with electrostatics included \cite{Nat14}.
Furthermore, $\chi$, as in the related lattice  regular solutions theories (e.g., the Flory-Huggins
theory \cite{Ter02}) describes the  short-range interactions between nearest neighbor (de)protonation
sites \cite{Avn20}. A parameter value $\alpha \leq 0$ encodes a favorable adsorption free energy,
while $\chi\geq0$ represents the tendency of particles on the macroion surface adsorption sites
to phase separate into domains. Fig.~\ref{Fig0} displays a schematic depiction of the charge
regulation process and the  Frumkin-Fowler-Guggenheim adsorption isotherm as a function of $\alpha$
for different values of the interaction parameter $\chi$.

\begin{figure*}[!t]
\centering
\includegraphics[width=17cm]{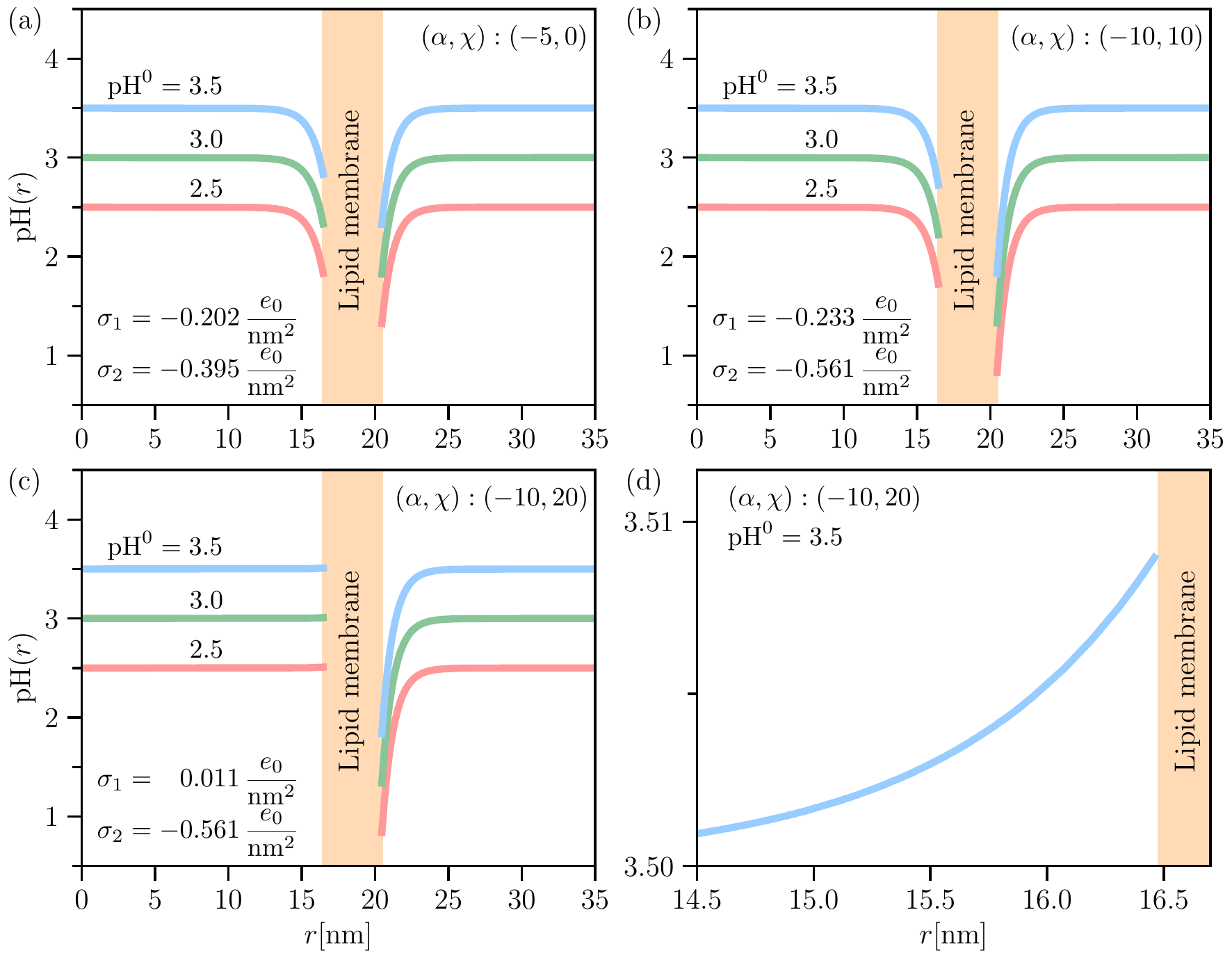}
\caption{Plot of $\textrm{pH}(r)$ across the membrane for different values of bulk $\textrm{pH}^0$ and $(\alpha, \chi) = (-5,0)$ [panel (a)], $(\alpha, \chi) = (-10,10)$ [panel (b)]. In both cases $\kappa_D = 1.215\, \textrm{nm}^{-1}$ and $R=16.46\, \textrm{nm}$ with dimensionless curvature $h$ fixed at $0.05$. The Bjerrum length $\ell_{B}=0.74\,\textrm{nm}$, $\epsilon_{p}=5$, $\epsilon_{w}=80$, surface dissociable group concentration $n_0=1\,\textrm{nm}^{-2}$, and $w=4 ~{\rm nm}$. $\sigma_{1}$ and $\sigma_{2}$ are obtained from the CR process. All curves show that the $\textrm{pH}$ vicinal to the bilayer reduces remarkably from $\textrm{pH}^0$ in the bulk. $\textrm{pH}$ at the outer surface is lower than that right at the inner surface and then increases exponentially towards $\textrm{pH}^0$ in the region far from the outer surface. (c)-(d) left panel $(\alpha, \chi) = (-10,20)$, right panel magnification for $\textrm{pH}$ close to the inner surface.  All curves  show that, $\textrm{pH}$ does not change much from $\textrm{pH}^0$. However, at the region close to the inner surface, we can observe change in $\textrm{pH}$ that grows as $\sinh (\kappa_D r)$. Again, the $\textrm{pH}$ goes exponentially towards $\textrm{pH}^0$ in the outer region. \label{Fig1}}
\end{figure*}

\begin{figure*}[!t]
\centering
\includegraphics[width=17cm]{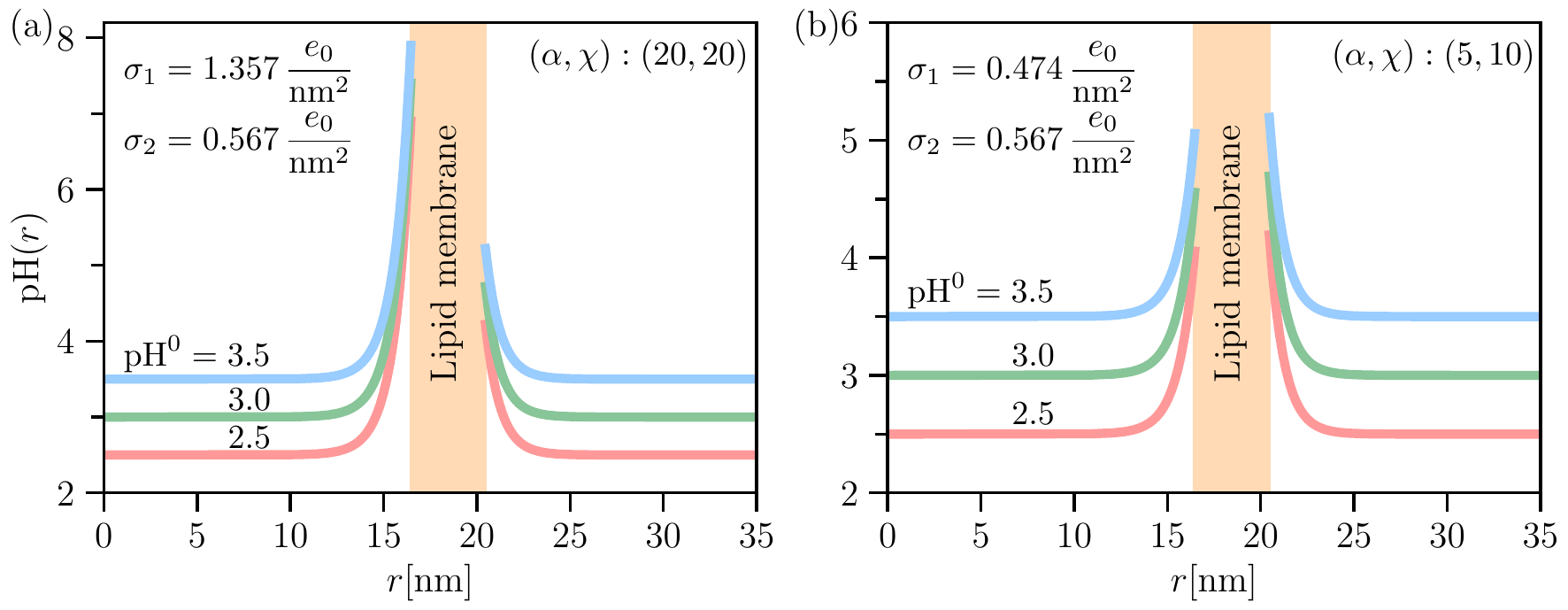}
\caption{\label{Fig3}{Plot of $\textrm{pH}(r)$ across the membrane. We have chosen $(\alpha,\chi) = (20,20)$ (a) and  $(\alpha,\chi) = (5,10)$ (b), $\kappa_D = 1.215 \, \textrm{nm}^{-1}$ and $h = (\kappa_D R)^{-1} = 0.05$. All curves show that, at the region close to the inner surface of the vesicle, $\textrm{pH}$ increases as $\sinh (\kappa_D r)$. The $\textrm{pH}$ decreases exponentially towards $\textrm{pH}^0$ in the region far from the outer surface for both panels (a) and (b). The Bjerrum length $\ell_{B}=0.74\,\textrm{nm}$, $\epsilon_{p}=5$, $\epsilon_{w}=80$, surface dissociable group concentration $n_0=1\,\textrm{nm}^{-2}$, and $w=4 ~{\rm nm}$.}}
\end{figure*}

It is important to reiterate at this point that other charge regulation models are of course possible and have
been proposed for various dissociable groups in different contexts \cite{Pod18,Bor01,Koo20}. Our reasoning
in choosing the particular Frumkin-Fowler-Guggenheim isotherm was guided by its simplicity in the way it takes
into account the salient features of the dissociation process on the membrane surface,  and the fact that the
implied phenomenology has been analyzed  before in the context of charged amphiphilic systems \cite{Har06}.

From the general electrostatic free energy Eq.~\ref{nkcajwgscfkj3} we then obtain the surface electrostatic
potential as
\begin{align}
\frac{\partial {\mathcal F}_{ES}(\sigma_1, \sigma_2, R)}{\partial \sigma_i} = 4\pi R_i^2~ \psi\left(\sigma_1, \sigma_2, R_i\right), 
\label{electrostatic1a}
\end{align}
for $i=1,2$. By considering Eqs.~\eqref{electrostatic1a}, as well as the form  of the charge regulation free energy
Eq.~\eqref{cr1} we derive the standard Frumkin-Fowler-Guggenheim adsorption isotherm \cite{Pod18,Koo20} from the
thermodynamic equilibrium obtained by minimizing the total free energy of the system. We get two equations that
correspond to charge regulation boundary conditions
\begin{equation}
\frac{\partial {\mathcal F}_{ES}(\sigma_1, \sigma_2, R)}{\partial \sigma_i} \frac{\partial \sigma_i}{\partial \eta_i} +  \frac{\partial {\mathcal F}_{CR}(\eta_i)}{\partial \eta_i} = 0 
\label{BC1a}
\end{equation}
for $i = 1,2$, which by taking into account Eq.~\eqref{electrostatic1a} can be solved by an implicit equation for
$\eta_i = \eta_i(\psi)$  \cite{Maj18, Maj19, Avn19, Avn20} in the form
\begin{align}
\eta_i(\psi) = \left({1  + e^{-\alpha - \chi \eta_i(\psi) + 2\beta e_0\psi}}\right)^{-1}
\label{BC2a}
\end{align}
with $\eta_i(\psi) = \eta_i(\psi(\sigma_1, \sigma_2, R_i))$. The numerical solution of the above equation is presented
in Fig.~\ref{Fig0} and corresponds to the Frumkin-Fowler-Guggenheim adsorption isotherm. Again we reiterate that the
DH linearizationpertains only to electrostatics, first term in Eq.~\ref{BC1a}, while the surface charging equilibrium,
is always considered in its full, non-linear form. It is evident from Fig.~\ref{Fig0} that for $\chi \leq -2\alpha$
the adsorption isotherm exhibits a discontinuous transition, whereas above the ``critical isotherm'', $\chi = - 2 \alpha$,
i.e., $\chi \geq -2\alpha$, it remain continuous.

\begin{table*}
\caption{$\textrm{pH}$ at the inner and outer surfaces of the vesicle. The bulk $\textrm{pH}^0$ is set as $5.2$. The radius of the vesicle is $100\,\textrm{nm}$. The Bjerrum length $\ell_{B}=0.74\,\textrm{nm}$, $\epsilon_{p}=5$, $\epsilon_{w}=80$, surface dissociable group concentration $n_0=1\,\textrm{nm}^{-2}$, and $w=4 ~{\rm nm}$. The inverse Debye length is $\kappa_D = 0.5\,\textrm{nm}^{-1}$.}
\renewcommand{\arraystretch}{1.3}
\centering
\resizebox{0.6\textwidth}{!}
{
\begin{tabular}{||c c c c c||}
 \hline
$(\alpha,\chi)$ & $\sigma_1 (e_0/\textrm{nm}^2)$ & $\sigma_2 (e_0/\textrm{nm}^2)$ & $\textrm{pH}(r=R)$ & $\textrm{pH}(r=R+w)$ \\ [0.5ex]
 \hline\hline
 $(\phantom{-}10,0\phantom{0})$ & $\phantom{-}0.232$ & $\phantom{-}0.280$ & $\phantom{0}7.003$ & $\phantom{0}7.268$ \\
 \hline
$(\phantom{-}10,20)$ & $\phantom{-}0.587$ & $\phantom{-}0.777$ & $\phantom{0}9.767$ & $10.930$ \\
 \hline
$(\phantom{-}20,20)$ & $\phantom{-}0.876$ & $\phantom{-}0.921$ & $11.961$ & $12.035$ \\
 \hline
$(\phantom{-}20,10)$ & $\phantom{-}0.646$ & $\phantom{-}0.797$ & $10.218$ & $11.091$ \\
 \hline
$(\phantom{\,\!}-5,10)$ & $0$ & $0$ & $\phantom{0}5.2\phantom{00}$ & $\phantom{0}5.2\phantom{00}$ \\
 \hline
$(-10,20)$ & $0$ & $0$ & $\phantom{0}5.2\phantom{00}$ & $\phantom{0}5.2\phantom{00}$ \\
 \hline
$(-10,10)$ & $-0.130$ & $-0.165$ & $\phantom{0}4.190$ & $\phantom{0}3.979$ \\
 \hline
$(-20,5\phantom{0})$ & $-0.429$ & $-0.526$ & $\phantom{0}1.871$ & $\phantom{0}1.310$ \\[1ex]
 \hline
\end{tabular}
}
\end{table*}

\begin{table*}
\caption{$\textrm{pH}$ at the inner and outer surfaces of the vesicle. The bulk $\textrm{pH}^0$ is set as $5.2$. The radius of the vesicle is $100\, \textrm{nm}$. The Bjerrum length $\ell_{B}=0.74\,\textrm{nm}$, $\epsilon_{p}=5$, $\epsilon_{w}=80$, surface dissociable group concentration $n_0=1\,\textrm{nm}^{-2}$, and $w=4 ~{\rm nm}$. The inverse Debye length is $\kappa_D = 1.0\,\textrm{nm}^{-1}$.}
\renewcommand{\arraystretch}{1.3}
\centering
\resizebox{0.6\textwidth}{!}
{
\begin{tabular}{||c c c c c||}
 \hline
$(\alpha,\chi)$ & $\sigma_1 (e_0/\textrm{nm}^2)$ & $\sigma_2 (e_0/\textrm{nm}^2)$ & $\textrm{pH}(r=R)$ & $\textrm{pH}(r=R+w)$ \\ [0.5ex]
 \hline\hline
 $(\phantom{-0}5,5\phantom{0})$ & $\phantom{-}0.376$ & $\phantom{-}0.452$ & 6.641 & $6.889$ \\
 \hline
$(\phantom{-}10,0\phantom{0})$ & $\phantom{-}0.449$ & $\phantom{-}0.515$ & $6.918$ & $7.128$ \\
 \hline
$(\phantom{-}10,20)$ & $\phantom{-}1.078$ & $\phantom{-}0.922$ & $9.307$ & $8.662$ \\
 \hline
$(\phantom{-}20,10)$ & $\phantom{-}1.078$ & $\phantom{-}0.922$ & $9.307$ & $8.662$ \\
 \hline
$(\phantom{\,\!}-5,10)$ & $0$ & $0$ & $5.2\phantom{00}$ & $5.2\phantom{00}$ \\
 \hline
$(-10,20)$ & $0$ & $0$ & $5.2\phantom{00}$ & $5.2\phantom{00}$ \\
 \hline
$(-10,10)$ & $-0.284$ & $-0.362$ & $4.111$ & $3.846$ \\
 \hline
$(-15,10)$ & $-0.563$ & $-0.697$ & $3.040$ & $2.595$ \\[1ex]
 \hline
\end{tabular}
}
\end{table*}

The boundary condition derived above, Eq. \ref{BC1a}, together with the solution of either full PB equation or the
linearized DH version for the electrostatic free energy Eq.~\eqref{fel} constitute the basic equations of our model.
In the case of the linear theory with the electrostatic free energy given by Eq.~\eqref{fel} we obtain the surface
potentials from Eq.~\eqref{electrostatic1a} as $\psi(R_i) = \psi_i (\eta_1, \eta_2)$.

Finally we should note that our approach is based on the free energy of the CR process and not on the assumed isotherms
that would follow from some chemical equilibrium considerations as is often done in the literature. While the two
approaches are in principle equivalent, it seems to us that the free energy approach has a more universal appeal and
allows also the explicit calculation of the total free energy, i.e., ES plus CR.

\section{Comparison between the full PB and the approximate DH solutions}

The DH approximation \cite{Muthu2023} is standardly invoked in order to derive limiting expressions and analytical
formulae in various contexts of macromolecular electrostatics  \cite{Roccia2021,Roccia2022}. In order to be able to
substantiate our usage of the DH approximation for most of the numerical results, we compare the $\textrm{pH}$ profile
resulting from the full PB equation with the consequences of the linearized DH equation. Again, we point out that the
linearization applies only to the electrostatic part, but not to the charge regulation part. Sometimes the linearization
is extended also to that case of charge regulation \cite{CARNIE1993260} as in the constant regulation boundary condition
often invoked by Borkovec \textit{et al.} \cite{BorkovecCR2004}, where for large separations one may expand the
charge-potential relationships at the surface around the potential at infinite separation.

We compare quantitatively the PB and the DH results for certain choices of the model parameters in Appendix \ref{appDHPB}.
The general conclusion  is that qualitatively, and often also quantitatively, they generally coincide but exhibit differences
in certain parts of the parameter space. It seems that one can thus safely use the DH approximation if the focus is on the
qualitative features, whereas a PB based calculation would be needed in order to do quantitative comparisons.

\section{Interfacial and luminal {pH}}

The above derivation and specifically the definition $\alpha = (\textrm{p}K_{\textrm{a}}-\textrm{pH}) \ln 10$ assume that
the concentration of the protons in solution is much lower then the concentration of salt and does not contribute to the
spatial profile of the electrostatic potential, neither on the PB nor the DH level. For many dissociable moieties at
physiological solution conditions this assumption holds well, but in general a more detailed implementation of the
$\textrm{pH}$ effects is needed, see e.g., \cite{Holm1,Holm2}.

With the above provisos the local $\textrm{pH}$ is a ``passive'' variable in the solution, except at the surface of the
lipid bilayer where it determines the dissociation state. The spatial dependence of $\textrm{pH} = -\log_{10}{[\textrm{H}^{+}]}$
is obtained from the electrostatic potential as
\begin{align}
\textrm{pH}(r) &= - \log_{10}{[\textrm{H}^+](r)} \notag\\
&= - \log_{10}{[\textrm{H}^+]_0} + \beta e_0 \psi(r) \log_{10}{e}\notag\\
&= \textrm{pH}^0 + \beta e_0 \psi(r) \log_{10}{e},
\label{pHdef}
\end{align}
where $\textrm{pH}^0= - \log_{10}{[H^+]_0}$ is either the $\textrm{pH}$ in the outside bulk reservoir, or the inside $\textrm{pH}$
that is set by the procedure of vesicle preparation and can coincide or can be different form the bulk value \cite{Chou1997}.
The calculation of $\textrm{pH}(r)$ then follows from the electrostatic potential profile that is written explicitly in the DH
approximation in Appendix \ref{DHapp}, yielding the $\textrm{pH}$ profile in the interior of the vesicle as
\begin{align}
\textrm{pH}(r\leq R) = - \log_{10}{[\textrm{H}^+]_0} + \beta e_0 ~\log_{10}{e} ~A \frac{\sinh(\kappa_D r)}{r},
\end{align}
where $\kappa_D$ is the inverse Debye length and $A$ is given in the Appendix \ref{DHapp}. Outside the vesicle the relevant
dependence is obtained as
\begin{align}
\textrm{pH}(r\geq R) = - \log_{10}{[\textrm{H}^+]_0} + \beta e_0 ~\log_{10}{e} ~B \frac{\exp (-\kappa_D r)}{r}
\end{align}
with $B$ given in the Appendix \ref{DHapp}. Obviously both inside as well as outside the vesicle the local $\textrm{pH}$ decays with
the Debye length. The two constants $A = A(\sigma_1, \sigma_2, \kappa_D w, \kappa_D R)$ and $B(\sigma_1, \sigma_2, \kappa_D w, \kappa_D R)$
are linear functions of the internal and external surface membrane charge densities. The above formulae allow us to explicitly obtain the
spatial variation of $\textrm{pH}$ around and across the membrane as well as the drops in $\textrm{pH}$ either across the membrane, or
between the bulk reservoir and the region adjacent to the membrane.

Note that while the electrostatic potential is well defined also inside the membrane, the $\textrm{pH}$ is not since the protons do not
move freely across the hydrophobic kernel of the membrane.

Because the $\textrm{pH}$ exhibits a spatially varying profile, $\textrm{pH} = \textrm{pH}(r)$, we can define different characteristic
values that can be,  at least in some cases, obtained either  directly or indirectly from  experiments \cite{Sarkar2018,Majumder2017}.
First we can define a drop in $\textrm{pH}$ across the membrane of the vesicle of magnitude
\begin{eqnarray}
\Delta \textrm{pH}_{\textrm{m}} =  \textrm{pH}(R+w) - \textrm{pH}(R). 
\label{DpHm}
\end{eqnarray}
Two other important quantifiers are the difference between the bulk $\textrm{pH}^0$ and $\textrm{pH}$ right at the outer surface as a
function of the outer radius of the vesicle $R$, defined as
\begin{eqnarray}
\Delta \textrm{pH}_{\textrm{out}} = \textrm{pH}(R + w) - \textrm{pH}^0. 
\label{DpHout}
\end{eqnarray}
Similarly, the difference between the bulk value $\textrm{pH}^0$ and $\textrm{pH}$ right at the inner surface as a function of the inner
radius of the vesicle $R$, defined as
\begin{eqnarray}
\Delta \textrm{pH}_{\textrm{in}} = \textrm{pH}(R) - \textrm{pH}^0.
\label{DpHin}
\end{eqnarray}
In what follows we will present several notable numeric results while at the same time reminding the reader that this is a multi-parameter
system and its parameter space cannot be explored exhaustively and systematically at this point.

A separate question here is the value of the potential, or equivalently the $\textrm{pH}$, at the center of the vesicle relevant for the
analysis of the lumina of viruses and virus-like particles \cite{Schoot}. Here again, we invoke the differences stemming from the different
procedures of preparation that can constrain the value of the inner $\textrm{pH}$ to be different from the bulk, a situation we will not
analyze in detail. In the case of the DH, small curvature approximation this is found to be
\begin{widetext}
\begin{align}
\lim_{r\longrightarrow 0}\psi(r) = \psi(0) =
\frac{ \epsilon_{w}\sigma_{1}R^{2}\kappa_D w^{2} + \epsilon_{p}[\sigma_{1}R^{2} + \sigma_{2}(R + w)^{2}]R +\epsilon_{w}\sigma_{1}(1 + \kappa_D R)w R^{2}}{\epsilon_{0}\epsilon_{w}\{(\epsilon_{p} - \epsilon_{w}) w^{2} + \epsilon_{p} R^{2} \}} \text{csch} (\kappa_D R),
\end{align}
\end{widetext}
taking into account the Appendix \ref{DHapp}. The $\textrm{pH}$ in the lumen then follows as
\begin{align}
\textrm{pH}(0) = - \log_{10}{[\textrm{H}^+]_0} + \beta e_0 ~\log_{10}{e} ~\psi(0),
\end{align}
where $\log_{10}{[\textrm{H}^+]_0}$ is the acidity in the bulk reservoir.  The expression for $\textrm{pH}(0)$ has two well defined limits
defined by
\begin{align}
\lim_{\kappa_D R \ll 1}\psi(0) = \frac{ \epsilon_{p}[\sigma_{1}R^{2} + \sigma_{2}(R + w)^{2}] +\epsilon_{w}\sigma_{1}w R}{\epsilon_{0}\epsilon_{w}\{(\epsilon_{p} - \epsilon_{w})\kappa_D w^{2} + \epsilon_{p}\kappa_D R^{2} \}},
\end{align}
and
\begin{align}
\lim_{\kappa_D R \gg 1}\psi(0) =&
\frac{ \epsilon_{w}\sigma_{1}R \left(R + w\right) 2 (\kappa_D R)(\kappa_D w) e^{-\kappa_D R} }{\epsilon_{0}\epsilon_{w}\{(\epsilon_{p} - \epsilon_{w})\kappa_D w^{2} + \epsilon_{p}\kappa_D R^{2} \}}.
\end{align}
Clearly in the second limit of $\kappa_D R \gg 1$ the potential in the center of the vesicle vanishes and thus the luminal $\textrm{pH}$
is the same as in the bulk, if the membrane is fully permeable to all mobile charged species. Interestingly enough, as can be discerned
from numerical solution, the electrostatic potential and consequently the $\textrm{pH}$ inside the vesicle are almost constant up to the
inner surface, implying that the Donnan potential approximation could be used for that case \cite{Schoot}. These are the only analytical
limits that one can derive for this problem.

Finally, we note that our calculation is based on the chemical equilibrium and ionic identity between the ionic solution inside and
outside the vesicle.

\section{Results}

We now analyze some numerical results obtained mostly within our model (1) unless specifically annotated for model (2).

We first investigate the full spatial profile of $\textrm{pH}$ in the vicinity of the vesicle. We assume that in the preparation of the
vesicle the inner and the outer solution are equilibrated at the same bathing solution $\textrm{pH}$. On the DH level the solution for
the potential can be derived  analytically, see Appendix \ref{DHapp}, but the solution of the CR isotherm, Eq.~\eqref{BC1a}, can only be
obtained numerically. The latter then yields the two surface charge densities, $\sigma_{1,2}$. Figures~\ref{Fig1}-\ref{Fig3} show the plots
of $\textrm{pH}$ vs $r$ obtained from Eq.~\eqref{pHdef} for different values of the bulk $\textrm{pH}^0$ and the CR parameters $\alpha, \chi$.
Clearly the general dependence of $\textrm{pH}(r)$ indicates a large variation close to both surfaces of the bilayer, to be quantified below.

We have used typical system parameters such as an inverse Debye length $\kappa_D=1.215\,\rm {nm}^{-1}$, or equivalently, screening length
$\lambda_D=0.823\,\textrm{nm}$ corresponding to an aqueous electrolyte solution with $140\,\textrm{mM}$ salt concentration. The dimensionless
curvature $h$ is defined as $h=1/(\kappa_{D}R)$, where $R$ is the inner radius of the vesicle. In Figs.~\ref{Fig1} and \ref{Fig3}, the
dimensionless curvature $h$ is fixed at $0.05$ corresponding to $R = 16.46\, \textrm{nm}$.

\begin{figure*}[!t]
\centering
\includegraphics[width=16cm]{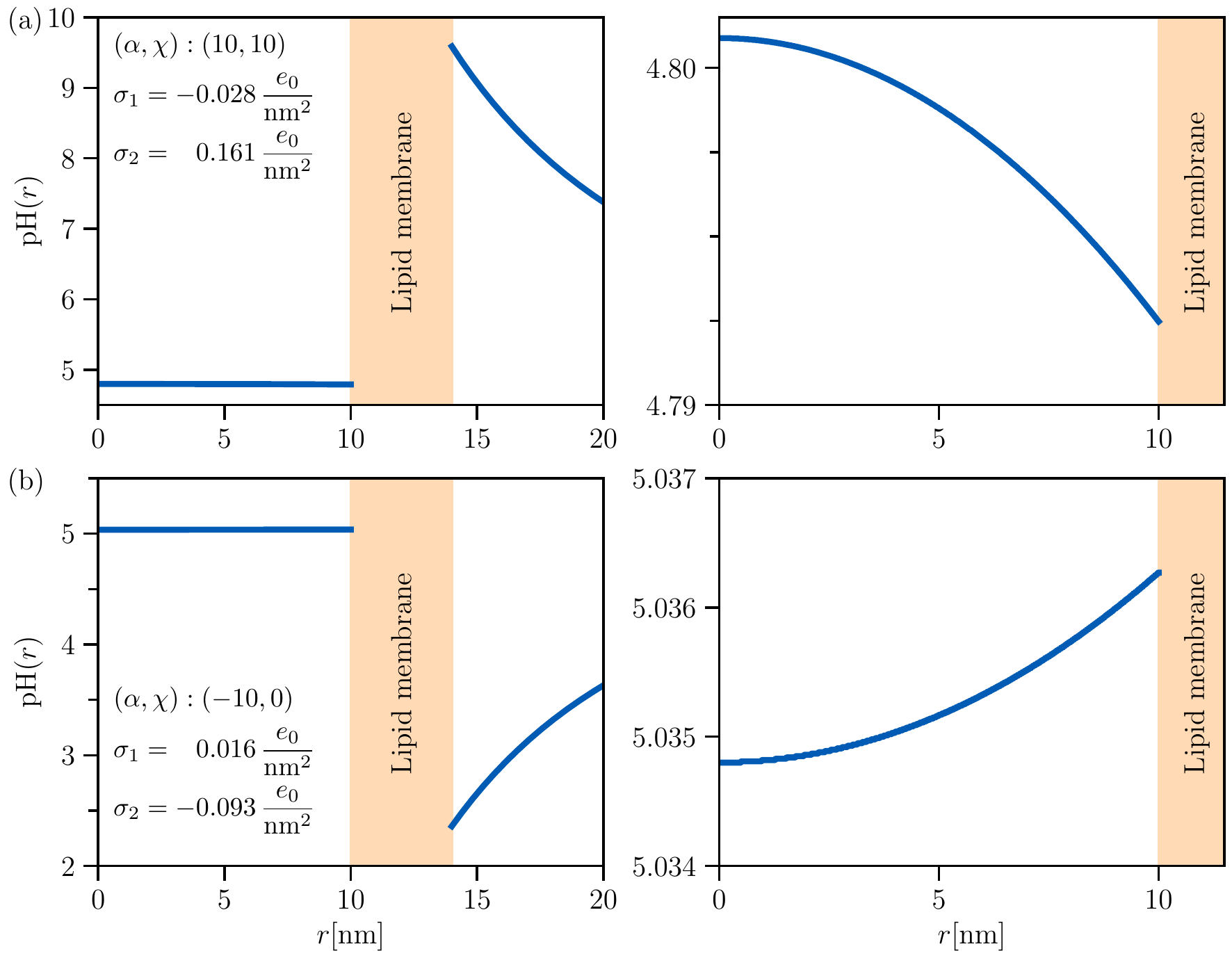}
\caption{Plot of $\textrm{pH}(r)$ across the membrane for bulk $\textrm{pH}^0 = 5.0$, panel (a) $(\alpha, \chi) = (10,10)$, panel (b) $(\alpha, \chi) = (-10,0)$. The right most panels show the expanded $\textrm{pH}(r)$ scale in order to see the small changes with $r$. In both cases $\kappa_D = 1/20 \, \textrm{nm}^{-1}$ and $R = 10 \, \textrm{nm}$  with dimensionless curvature $h$ fixed at $2$.  $\sigma_{1}$ and $\sigma_{2}$ are obtained from the CR process and correspond to the values of $(\alpha, \chi)$ combination. The vicinal $\textrm{pH}$ close to the outer surface of the vesicle can be drastically different from the bulk one, depending on the parameters. The Bjerrum length $\ell_{B}=0.74\,\textrm{nm}$, $\epsilon_{p}=5$, $\epsilon_{w}=80$, surface dissociable group concentration $n_0=1\,\textrm{nm}^{-2}$, and $w=4 ~{\rm nm}$. \label{Fig1a}}
\end{figure*}

\begin{figure*}[!t]
\centering
\includegraphics[width=17cm]{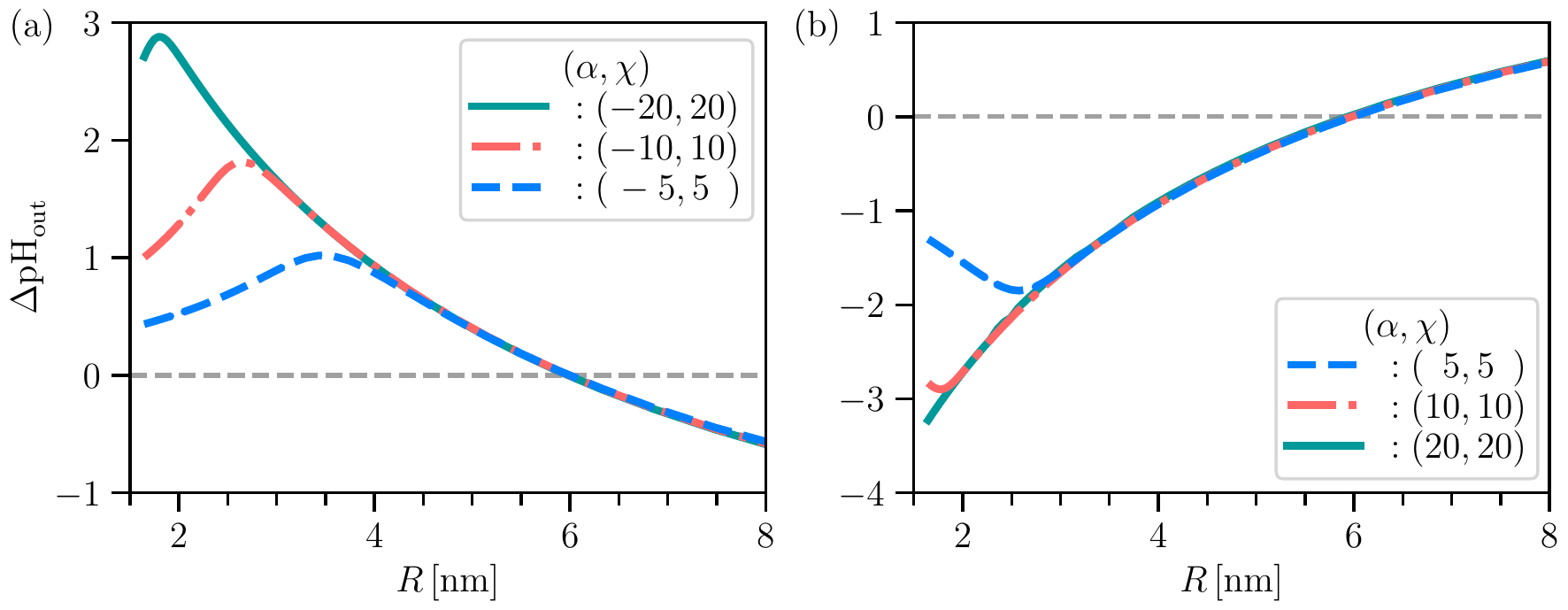}
\caption{{Plot of $\Delta\textrm{pH}_{\text{out}}$ as defined in Eq.~\eqref{DpHout} vs $R$ with $\kappa_D = 1.215 \, \textrm{nm}^{-1}$. Clearly the $\textrm{pH}$ vicinal but exterior to the vesicle can be larger or smaller than the bulk $\textrm{pH}^{0}$, depending on the charge regulation parameters entering into the CR model. In general, $\alpha \leq 0$ makes the vicinal $\textrm{pH}$ larger, while $\alpha \geq 0$ makes it smaller. Also, the larger $\chi$ the larger is this effect.\label{Fig5}}}
\end{figure*}

\begin{figure*}[!t]
\centering
\includegraphics[width=17cm]{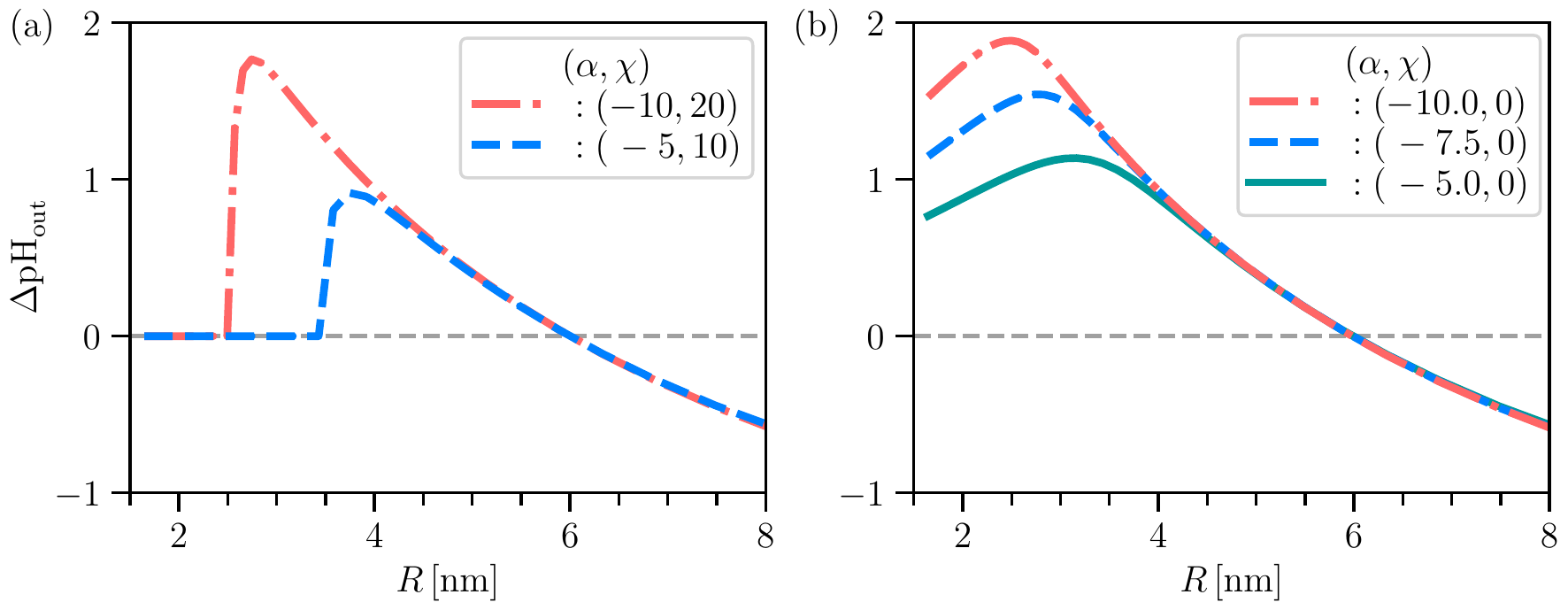}
\caption{\label{Fig6}{Plot of $\Delta\textrm{pH}_{\text{out}}$ vs $R$ for different values of $(\alpha, \chi)$ and $\kappa_D = 1.215 \, \textrm{nm}^{-1}$. (a) $\chi = -2\alpha$. (b) $\alpha < 0$ and $\chi = 0$ (no surface interaction). For large enough positive $\chi$ the dependence of $\Delta\textrm{pH}_{\text{out}}$ on $R$ shows a behavior akin to a second order transition, where for small radii it vanishes, and then at a critical value $R = R_c$ it starts deviating from zero, reaching for a maximum and then levelling off at a constant value for a sufficiently large $R$. No such behavior is observed for $\chi =0$. }}
\end{figure*}

From our Frumkin-Fowler-Guggenheim charge regulation model, we obtain $(\sigma_{1} = -0.202\, e_0/\textrm{nm}^2, \sigma_{2} = -0.395\, e_0/\textrm{nm}^2)$
for $(\alpha = -5, \chi = 0)$ (no surface interaction) (Fig.~\ref{Fig1}(a)) and $(\sigma_{1} = -0.233\, e_0/\textrm{nm}^2, \sigma_{2} = -0.561\, e_0/\textrm{nm}^2)$
for $(\alpha = -10, \chi = 10)$ (Fig.~\ref{Fig1}(b)). The plots show that $\textrm{pH}$ decreases remarkably at the region close to the vesicle's
inner surface and that the $\textrm{pH}$ at the outer surface is lower than that at the inner surface. In general the $\textrm{pH}$ approaches
exponentially the bulk $\textrm{pH}^0$ away from the membrane.  For negative $\alpha = -10$ and $\chi = 20$ ($\chi = -2\alpha$) case, (Fig.~\ref{Fig1}(c)),
the Frumkin-Fowler-Guggenheim charge regulation model yields $(\sigma_{1} = 0.011\, e_0/\textrm{nm}^2, \sigma_{2} = -0.561\, e_0/\textrm{nm}^2)$.
Close to the surface the $\textrm{pH}$ increases like $\sinh (\kappa_D r)$ which is shown with more detail in Fig.~\ref{Fig1}(d).

\begin{figure*}[!t]
\centering
\includegraphics[width=17cm]{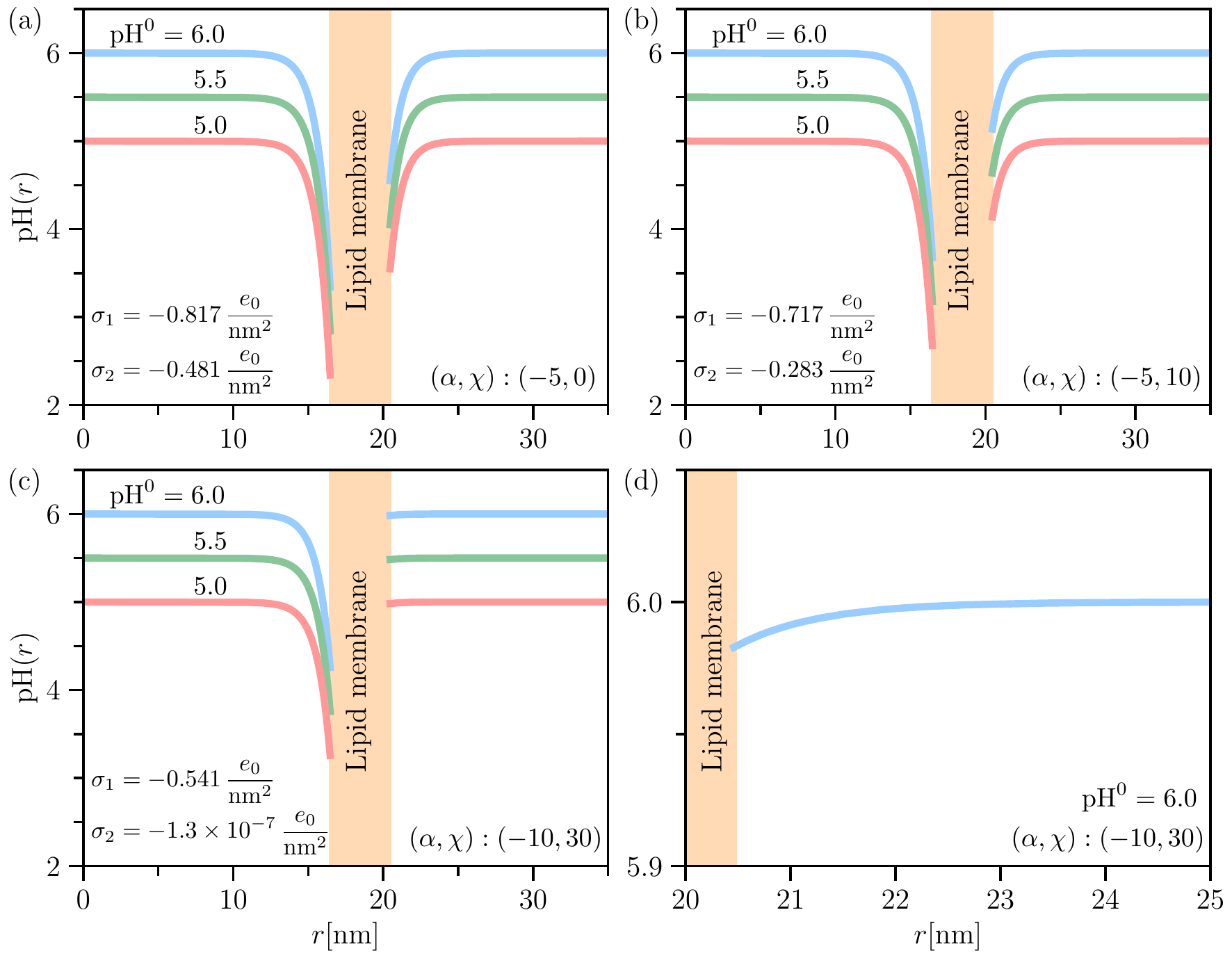}
\caption{Plot of $\textrm{pH}(r)$ across the membrane for model (2), Eq. \eqref{sigma2}, with $\kappa_D = 1.215~{\textrm{nm}}^{-1}$ and $R = 16.46$ nm with dimensionless curvature $h$ set at 0.05. The surface charge densities on both sides of the bilayer $\sigma_1$ and $\sigma_2$ are negative. For $(\alpha, \chi) = (-5, 0)$ or $(\alpha, \chi) = (-5, 10)$ the corresponding $\Delta \textrm{pH}_{\textrm{out}}$ is comparable. For $(\alpha, \chi) = (-10, 30)$, $\Delta \textrm{pH}_{\textrm{out}}$ is much smaller and the deviation of $\textrm{pH}$ from $\textrm{pH}^0$ at the outer surface can only be seen clearly after magnification close to the outer surface (d).
\label{Fig7}}
\end{figure*}

For positive $\alpha > 0$, Fig. \ref{Fig3}, we obtain  $(\sigma_{1} = 1.357\, e_0/\textrm{nm}^2, \sigma_{2} = 0.567\, e_0/\textrm{nm}^2)$ at $(\alpha = 20, \chi = 20)$ (Fig.~\ref{Fig3}(a))
and $(\sigma_{1} = 0.474\, e_0/\textrm{nm}^2, \sigma_{2} = 0.567\, e_0/\textrm{nm}^2)$ at $(\alpha = 5, \chi = 10)$
(Fig.~\ref{Fig3}(b)). All curves exhibit the same scaling  $\sinh (\kappa_D r)$ near the inner surface. Clearly in the case of $\alpha > 0$ the local  $\textrm{pH}$ decreases from the value it attains near the surface of the bilayer towards the bulk $\textrm{pH}^0$ for both Figs.~\ref{Fig3}(a) and~\ref{Fig3}(b).

{In order to see the deviation of $\textrm{pH}$ from the bulk $\textrm{pH}^0$ clearly, we have used 
$\kappa_D = 1/20 \, \textrm{nm}^{-1}$ and $R = 10 \, \textrm{nm}$ with dimensionless curvature $h$ fixed at $2$. Figure~\ref{Fig1a} presents the plot of $\textrm{pH}(r)$ across the membrane for bulk $\textrm{pH}^0 = 5.0$, $(\alpha, \chi) = (10,10)$ (Fig.~\ref{Fig1a}(a)), $(\alpha, \chi) = (-10,0)$ (Fig.~\ref{Fig1a}(b)). The right most panels show the expanded $\textrm{pH}(r)$ scale in order to see the small changes with $r$.  The vicinal $\textrm{pH}$ close to the outer surface of the vesicle can be drastically different from the bulk one, depending on the parameters.}

We now analyze the dependence of the $\Delta\textrm{pH}_{\textrm{out}}$  on the various parameters of the system in Figs. \ref{Fig5} and \ref{Fig6}. The procedure is again the same as before, we solve analytically for the electrostatic potential and then obtain the corresponding surface charges from the  Frumkin-Fowler-Guggenheim charge regulation model. {The inverse Debye length is set as $\kappa_D = 1.215 \, \textrm{nm}^{-1}$.} Of particular importance is the dependence on the curvature of the bilayer. In Figs. \ref{Fig5} and \ref{Fig6} we show this dependence for positive and negative $\alpha$. Clearly at first  $\Delta\textrm{pH}_{\textrm{out}}$ starts with a positive value, meaning that the surface $\textrm{pH}$ is larger then the bulk $\textrm{pH}^0$. It then increases with curvature, reaches a maximum and then decays, eventually even turning negative. This behavior is the more pronounced the more $\alpha$ is negative. For positive values of $\alpha$,  $\Delta\textrm{pH}_{\textrm{out}}$ starts with a negative value, reaches a minimum and then increases, eventually turning positive. Interestingly enough, because of the properties of the Frumkin-Fowler-Guggenheim isotherm, for small curvatures $\Delta\textrm{pH}_{\textrm{out}}$ can start at zero, see Fig. \ref{Fig6}(a), for certain negative values of $\alpha$. This simply means that at that $\alpha, h$ the bilayer is uncharged, charges up at a critical value of curvature and then follows basically the same behavior as for other negative values of $\alpha$. The pronounced variation of $\Delta\textrm{pH}_{\textrm{out}}$, which is in principle measurable \cite{Sarkar2018}, indicates that one could get some indication for the numerical values of the model parameters by comparing with suitable experiments.

\begin{figure*}[!t]
\centering
\includegraphics[width=16cm]{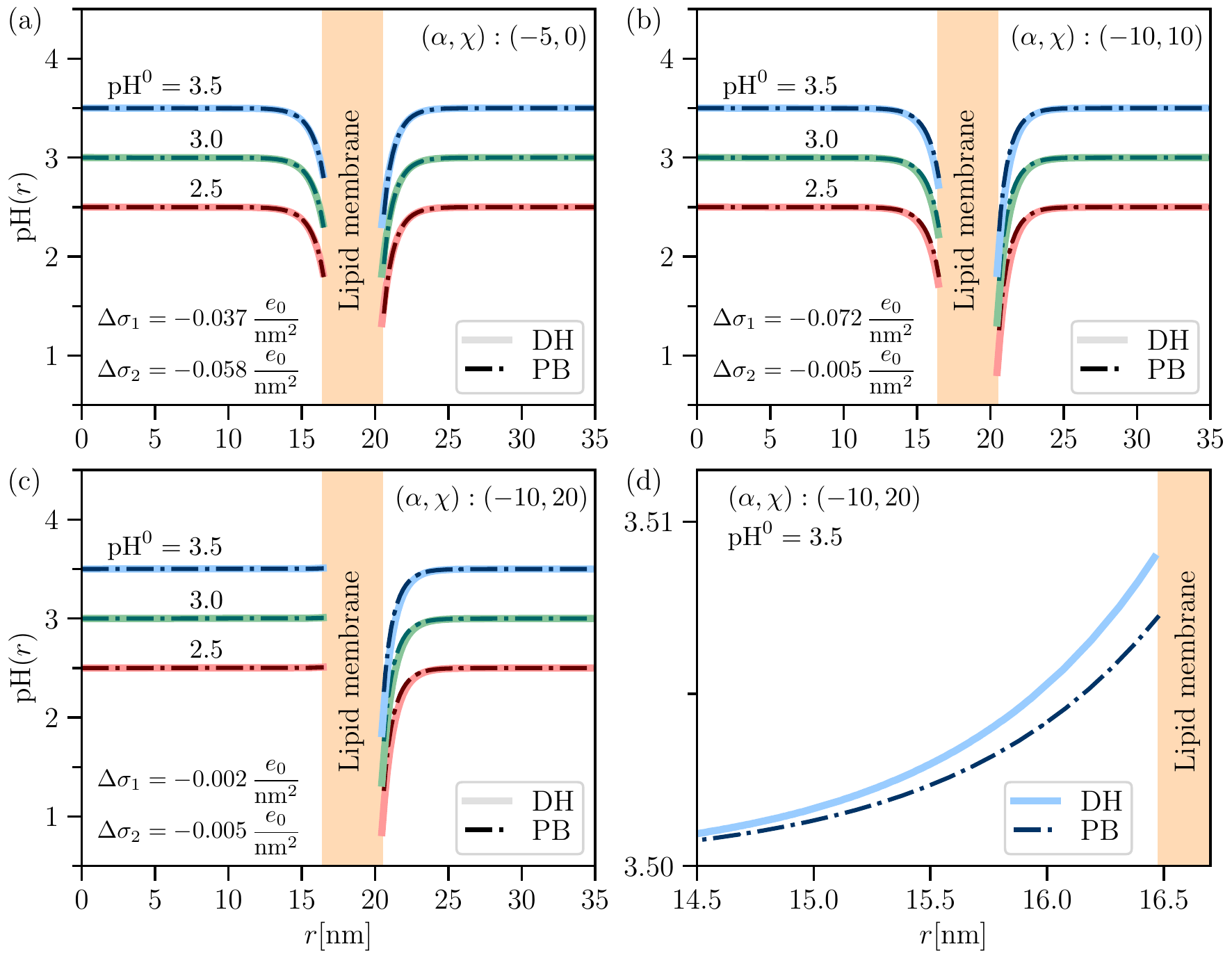}
\caption{\label{Fig8}{Comparison of the spatial dependence of $\textrm{pH}(r)$ across the membrane for different values of the bulk $\textrm{pH}^0$ obtained from the full PB and the linearized DH solutions in the case of $\alpha \leq 0$. (a) $(\alpha, \chi) = (-5, 0)$, (b) $(\alpha, \chi) = (-10, 10)$, (c) $(\alpha, \chi) = (-10, 20)$ for different values of the bulk $\textrm{pH}^0$ and (d) an expanded view of the region vicinal to the membrane for $\textrm{pH}^0 = 3.5$. Here  $\Delta\sigma_i = \sigma_i^{\text{PB}} - \sigma_i^{\text{DH}}$. 
The absolute value of $\sigma_i$ is greater within the full PB theory, but it changes when there is a symmetry breaking, i.e., for $(\alpha, \chi) = (-10, 20)$. $\Delta\sigma_1$ is negative here but $\sigma_1$ is positive in this case, implying that $ \sigma_1^{\text{PB}} < \sigma_1^{\text{DH}}$. The largest discrepancies are observed in the vicinity of the membrane but even then they are overall small.}}
\end{figure*}

\begin{figure*}[!t]
\centering
\includegraphics[width=16cm]{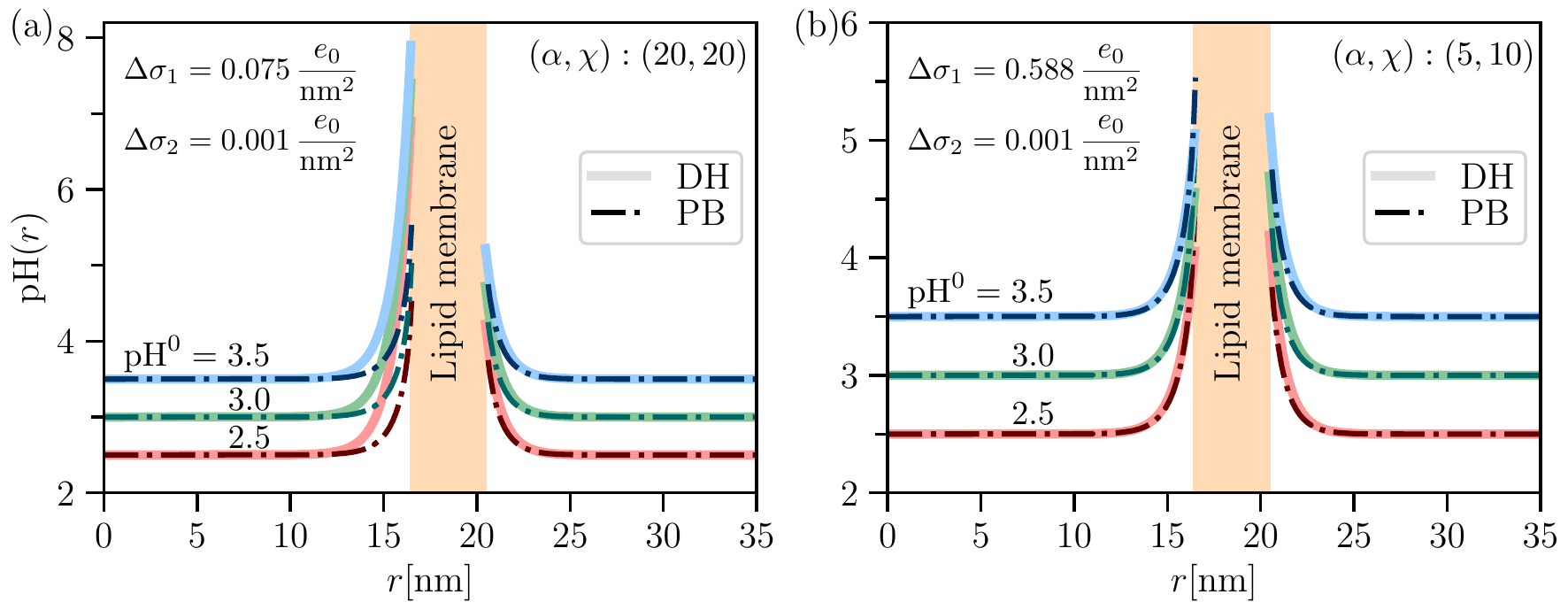}
\caption{\label{Fig9}{Comparison of the spatial dependence of $\textrm{pH}(r)$ across the membrane for different values of the bulk $\textrm{pH}^0$ obtained from the full PB and the linearized DH solutions in the case of $\alpha \geq 0$. (a) $(\alpha, \chi) = (20, 20)$ and (b) $(\alpha, \chi) = (5, 10)$. One observes a  significant deviation between $\textrm{pH}$ values obtained from PB and DH approaches for $(\alpha, \chi) = (20, 20)$. For $(\alpha, \chi) = (5, 10)$ the difference is nevertheless small though $\Delta\sigma_1$ is higher in this case. This is because of the value of $\sigma_1$ itself which is high for $(\alpha, \chi) = (20, 20)$.}}
\end{figure*}

\begin{figure}[!t]
\centering
\includegraphics[width=7cm]{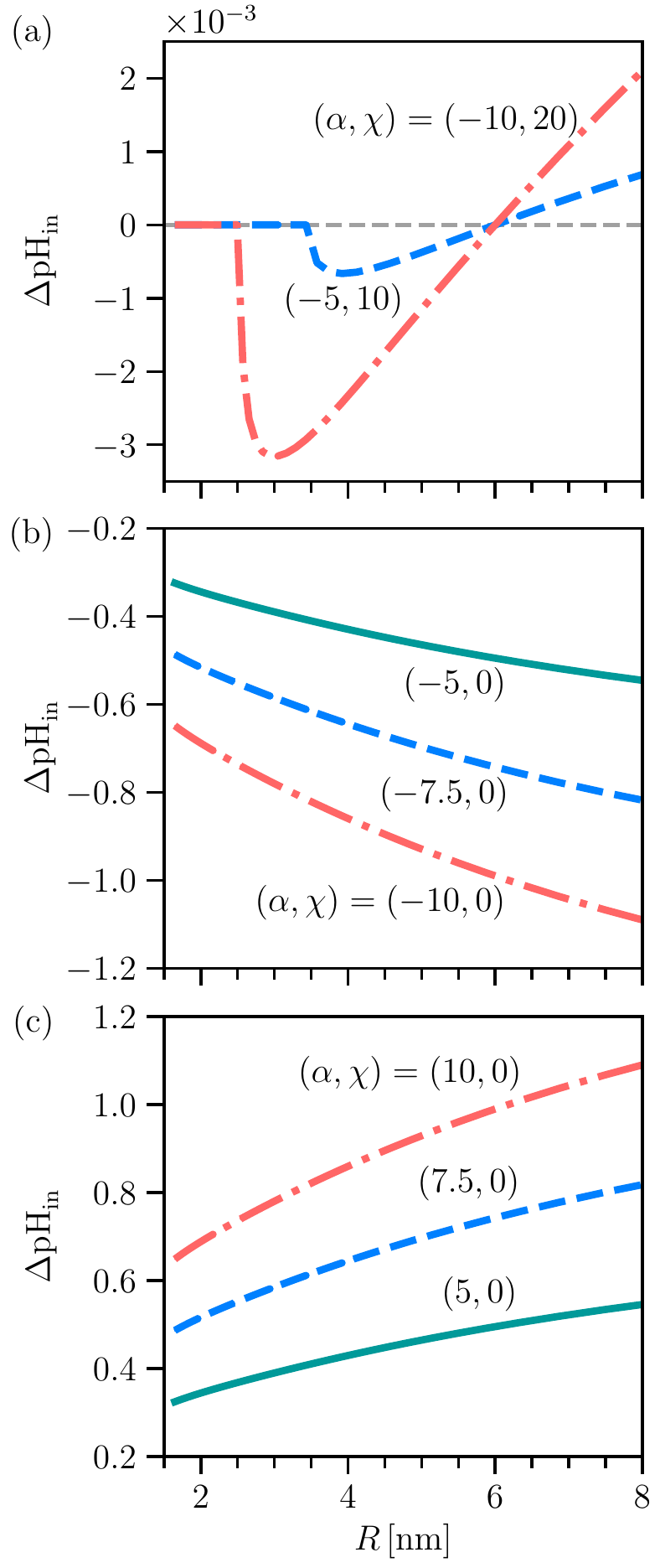}
\caption{\label{Fig10}{Plot of $\Delta\textrm{pH}_{\textrm{in}}$ vs $R$ for the "critical isotherm" $\chi = -2\alpha$ case (a) and for $\chi = 0$ for negative (b) and positive (c) values of $\alpha$. Overall $\Delta\textrm{pH}_{\textrm{in}}$ is very small with different behaviors in a curve i.e., (i) $\Delta\textrm{pH}_{\textrm{in}} < 0$ (ii)  $\Delta\textrm{pH}_{\textrm{in}} = 0$, and (iii) $\Delta\textrm{pH}_{\textrm{in}} > 0$, i.e. $\Delta\textrm{pH}_{\textrm{in}}$ can increase, decrease or remain constant as the radius $R$ varies.}  {(b,c) Plots of $\Delta\textrm{pH}_{\textrm{in}}$ vs $R$ for $\chi = 0$ case, that corresponds to no surface interaction. Plots are for $\alpha < 0$ (b) and $\alpha > 0$ (c). The $\Delta\textrm{pH}_{\textrm{in}}$ in both cases are in fact the same in absolute value but different in sign.} For all cases $\kappa_D = 1.215\,\textrm{nm}^{-1}$.}
\end{figure}

In a recent detailed experimental work on detection of curvature-dependent interfacial $\textrm{pH}$ for amphiphilic self-assemblies and unilamellar phospholipid vesicles an interface-interacting spiro-rhodamine $\textrm{pH}$ probe and Schiff base polarity probe have been used to measure the deviation of the local interfacial $\textrm{pH}$ from the bulk phase \cite{Sarkar2018}. It has been shown that the charging state (and polarity) of the 
amphiphile and phospholipid self-assemblies can be regulated by the curvature of the vesicle/micelle. While the experimental system is more complicated than our model and specifically contains also the interfacial dielectric constant 
we believe it could be instructive to compare the predictions of our model with the measured values for $\Delta\textrm{pH}_{\textrm{out}}$. 

We calculated $\textrm{pH}(r = R)$ and $\textrm{pH}(r = R + w)$ from the Frumkin-Fowler-Guggenheim CR model as described in detail above. The curvature radius $R$ is set as $100\,\textrm{nm}$ which is one of the experimentally chosen values for the large unilamellar phospholipid vesicles in the experiment \cite{Sarkar2018}. Other examples include radii $  \sim 15, 25, 50$ nm that we did not consider explicitly. We assumed the bulk value $\textrm{pH}^0 = 5.2$ corresponding to the $2.0\,\textrm{mM}$ cacodylate-HCl buffer. The inverse ionic screening length was taken as $\kappa_D = 0.5\,\textrm{nm}^{-1}$ (Table 1.) and $\kappa_D  = 1.0 ~\textrm{nm}^{-1}$ (Table 2.), corresponding to ionic concentrations of $25\,\textrm{mM}$ and $100\,\textrm{mM}$. For both choices $\textrm{pH}(r = R)$ {and $\textrm{pH}(r = R + w)$ are} larger than the bulk $\textrm{pH}^0$ for $\alpha > 0$ and less than the bulk $\textrm{pH}^0$ for $\alpha < 0$. In addition for the critical adsorption isotherm $\chi = -2 \alpha$ we have {$\textrm{pH}(r = R) = \textrm{pH}(r = R + w) = \textrm{pH}^0$} for small radii of curvature. 

In experiments of \cite{Sarkar2018} performed for 1,2-dimyristoyl-sn-glycero-3-phosphorylglycerol (DMPG)/1,2-dimyristoyl-sn-glycero-3-phosphocholine (DMPC) $(2:1)$ mixture in the case of large unilamellar vesicles, the authors obtained $\Delta \textrm{pH}_{\textrm{out}} \simeq -1.4 - -1.6$. In this particular case the negative charge stems from the deprotonation of DMPG while the DMPC lipid component carries no net charge. {We model this situation both in the framework of model (1),  as well as model (2),   which seems to be the more realistic case. From our calculations in the framework of the model (1), with $(\alpha, \chi) = (-10, 10)$, we obtain $\Delta \textrm{pH}_{\textrm{out}}$ (defined in Eq.~\eqref{DpHout}) as $-1.221$ (Table 1.) and $-1.354$ (Table 2.). Theoretically, depending on the combination of $(\alpha, \chi)$, the value of $\Delta \textrm{pH}_{\textrm{out}}$ from our model 1 could be obtained in the range of the experiments.} 

On the other hand the model (2), Fig. \ref{Fig7}, yielding only negative values of the surface charge density and thus, as already stated, being closer to the experimental situation  corresponding to  DMPG deprotonation,  with $(\alpha, \chi) = (-5, 0)$ and $(\alpha, \chi) = (-5, 10)$, yields $\Delta \textrm{pH}_{\textrm{out}}$ up to -1.5, which is again close to the stated experimental value \cite{Sarkar2018}. Note also that in model (2) the corresponding CR  parameters $(\alpha, \chi)$ corresponding to a close match with  experimental data are much smaller and thus possibly more realistic.
 
 \section{Discussion}

The properties of the bulk bathing solution can be quite different from the local  environment near the embedded membrane proteins and lipid  functional groups. This is in particular important for the spatial dependence of the  acidity/basicity that actually governs the dissociation equilibrium of protein and lipid dissociable groups. Recent detailed experiments \cite{Majumder2017,Sarkar2018} have in fact shown that the surface $\textrm{pH}$ can differ from the bulk one by several units. A deviation of $1.8$ and $2.2$ units from the bulk to the interface was detected for cationic cetrimonium bromide (CTAB) micelles and dimethyldioctadecylammonium bromide (DDAB) unilamellar vesicles \cite{Majumder2017}, and the $\textrm{pH}$ deviation of $-1.4$ to  $-1.6$ for  DMPG/DMPC $(2:1)$ mixture large unilamellar vesicles \cite{Sarkar2018}, respectively. 

Motivated by these experimental findings we performed a detailed theoretical study of the effects of interfacial curvature in the case of a bilayer vesicles with surface dissociable groups, on the interfacial acidity/basicity properties and found that the local $\textrm{pH}$ can actually veer off quite far from the nominal values set in the bulk. Our theoretical model is based on the Poisson-Boltzmann volume free energy functional for mobile charges and the Frumkin-Fowler-Guggenheim  adsorption isotherm model, formulated in terms of the appropriate surface free energy, for the surface  dissociation  equilibrium. It quantifies the surface adsorption/dissociation  energy, the nearest neighbor surface interaction energy and the lattice gas entropy of adsorption/dissociation sites. In general the model is rich enough to encode the simple Langmuir isotherm behavior captured also by the original charge regulation model \cite{Nin71} as well as the more nuanced behavior of the Frumkin-Fowler-Guggenheim  isotherm \cite{Maj18} and the ensuing first order adsorption/dissociation transition, see Fig. \ref{Fig0}. We solved the model numerically on the full non-linear PB level as well as in the linearized DH approximation for a spherical vesicle with finite thickness permeable membrane,  whose solvent accessible surfaces contains the dissociable moieties. We were specifically interested to derive the spatial profiles of the vicinal as well as luminal $\textrm{pH}$ as a function of the parameters of the model. 

\begin{figure*}[!t]
\centering
\includegraphics[width=17cm]{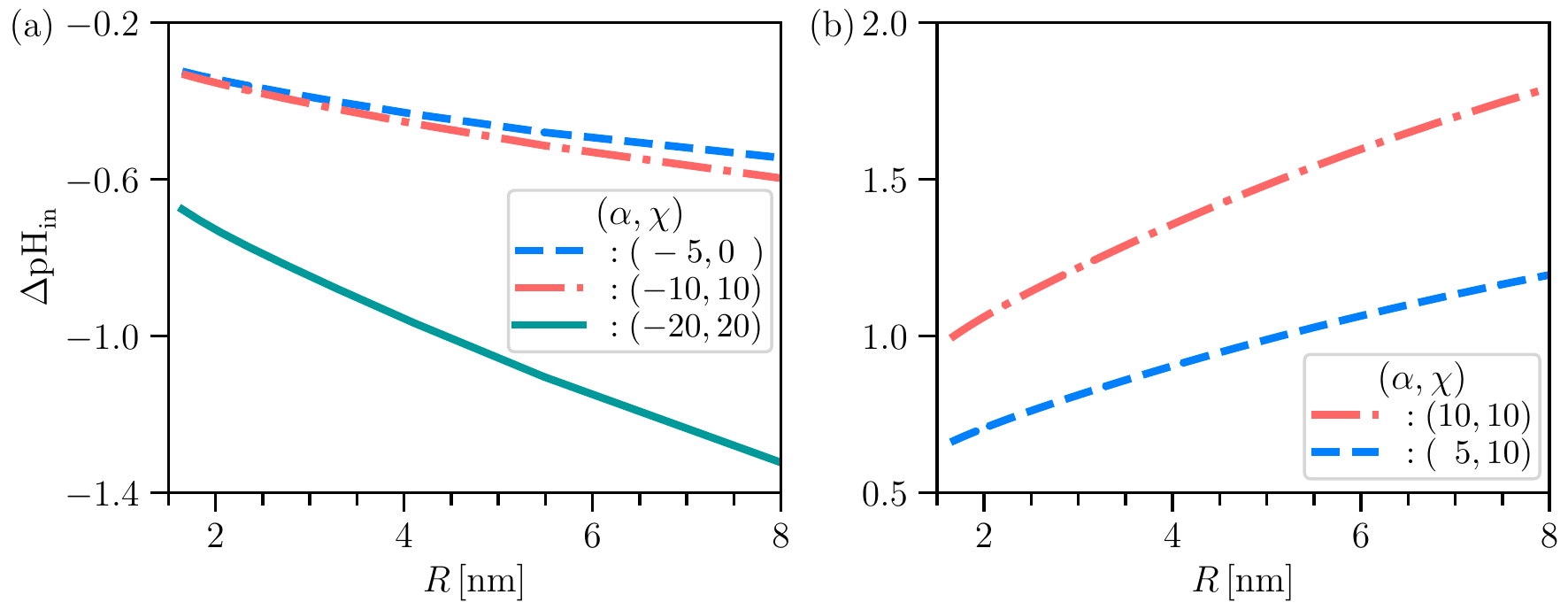}
\caption{\label{Fig4gfw}{Plot of $\Delta\textrm{pH}_{\textrm{in}}$ vs $R$. The inverse Debye length $\kappa_D$ is fixed at $1.215\,\textrm{nm}^{-1}$. $\sigma_{1}$ and $\sigma_{2}$ are obtained from the CR process. The plots are shown for $\alpha < 0$ (a) and $\alpha > 0$ (b). $\Delta\textrm{pH}_{\textrm{in}}$ decreases (a)/increases (b) when the size of the vesicle increases.}}
\end{figure*}

The numerical solutions of our model predict the full spatial dependence of $\textrm{pH}$ as a function of surface dissociation model parameters, $\alpha$ and $\chi$, assumed to be the same for the outer (solution interface) as well as inner (luminal interface) bilayer surface. What is clear is that for $\alpha \leq 0$ the $\textrm{pH}$ vicinal to the bilayer differs remarkably from $\textrm{pH}^0$ of the bulk, with $\textrm{pH}$ at the outer surface being lower than that right at the inner surface, see Fig. \ref{Fig1}. This behavior can be modified by the amount of salt in the bulk reservoir, or equivalently its screening length ${\kappa_{D}}^{-1}$, which can be seen by comparing the behavior of Figs. \ref{Fig1} and \ref{Fig1a}.  As for the case of $\alpha \geq 0$, see Fig. \ref{Fig3}, clearly the $\textrm{pH}$ at the luminal interface is much more perturbed than at the solution interface, implying that the negative curvature has quantitatively a larger effect than the positive curvature of the interface. 

We also compared two different models of CR process, one associated with a symmetric charge distribution across the bilayer (model 1), and another one with an asymmetric distribution (model 2). We note that the $\textrm{pH}$ profiles obtained can be similar, the CR parameters $(\alpha, \chi)$ corresponding to that profile are quite different, compare Figs.~\ref{Fig1} and \ref{Fig7}.

The change in $\textrm{pH}$ at the solution interface can be quantified further by computing $\Delta\textrm{pH}_{\text{out}}$ as defined in Eq.~\eqref{DpHout}. From Fig. \ref{Fig5} we discern that for $\alpha \leq 0$, $\Delta\textrm{pH}_{\text{out}}$ as a function of the membrane curvature develops a local maximum whose position depends on the radius of the vesicle, $R$, contrary to the case of $\alpha \geq 0$ where $\Delta\textrm{pH}_{\text{out}}$ develops a local minimum as a function of the radius of the vesicle. In both cases the position of the extremum depends also on the value of the interaction parameter $\chi$ displacing it towards the interface for larger positive values. The value of  $\Delta\textrm{pH}_{\text{out}}$ can be either positive or negative depending on the surface interaction parameters. For some combinations of $(\alpha, \chi)$, $\Delta\textrm{pH}_{\text{out}}$ can show a second order transition as a function of $R$, being zero for small enough $R$ and then starting to deviate from zero at a critical value of the radius of the vesicle, see Fig. \ref{Fig6}. This can only happen for large enough nearest neighbor interactions at the surface, $\chi \neq 0$.

Another notable conclusion, following from the comparison between the DH approximation and the full PB solution, is that the former describes the same effects as the latter on a qualitative level. Notably, the DH approximation as used here does not imply just a DH solution for the electrostatic potential but also a full minimization of the final free energy with non-linear surface interaction terms included. It would be probably more appropriate to refer to it as the DH-CR solution than as merely a DH solution. We surmize that it could be used to great advantage also in other situations where the full PB solution is prohibitively difficult to find even numerically.

\section*{Funding}

PK, RH and RP acknowledge funding from the Key Project No. 12034019 of the National Natural Science Foundation of China and the 1000-Talents Program of the Chinese Foreign Experts Bureau.

\section*{Acknowledgements}
RP, PK and HR acknowledge the support of the School of Physics, University of Chinese Academy of Sciences, Beijing. RP also acknowledges the support of the Wenzhou Institute of the University of Chinese Academy of Sciences, Wenzhou, Zhejiang.

\section{Appendix}\label{sec:appendix}

\subsection{Debye-H\" uckel solution} \label{DHapp}

In the DH approximation,  the electrostatic potential can be obtained explicitly for different regions of the problem \cite{Siber2007}. Inside the vesicle,
\begin{align}
\frac{1}{r}\frac{d^2 (r\Phi_{\textrm{I}}(r)) }{dr^2} + {\kappa_D}^2 \Phi_{\textrm{I}}(r) = 0
\end{align}
giving
\begin{align}
\Phi_{\textrm{I}}(r\leq R) = A \frac{\sinh(\kappa_D r)}{r}.
\end{align}
In the lipid membrane or the  amphiphilic layer we have
\begin{align}
\frac{1}{r}\frac{d^2 (r\Phi_{\textrm{II}}(r)) }{dr^2} = 0
\end{align}
implying
\begin{align}
\Phi_{\textrm{II}}(R \leq r \leq R+w) = \frac{C}{r} + D,
\end{align}
while in the external compartment 
\begin{align}
\frac{1}{r}\frac{d^2 (r\Phi_{\textrm{III}}(r)) }{dr^2} + {\kappa_D}^2 \Phi_{\textrm{III}}(r) = 0
\end{align}
is satisfied by
\begin{align}
\Phi_{\textrm{III}}(r\geq R+w) = B \frac{\exp (-\kappa_D r)}{r}.
\end{align}
Here the inverse square of the Debye screening length is given by $\kappa_D^{2} = 2 n_I \beta e_0^2/\varepsilon_w\varepsilon_0$, where $n_I$ is the  bulk electrolyte ionic concentration. 

Above we have obviously assumed that the two compartment (I and III) are in chemical equilibrium and can exchange electrolyte solution ions. If this is not the case, the screening properties, $\kappa_D(\textrm{I})$ and $\kappa_D(\textrm{III})$ would differ.

The electrostatic potentials in different regions are connected via boundary conditions that have the standard form
\begin{equation}
\left. \epsilon_w \frac{\partial \Phi_{\textrm{I}}(r)}{\partial r} \right |_{r = R} - 
\left. \epsilon_p \frac{\partial \Phi_{\textrm{II}}(r)}{\partial r} \right |_{r = R} = \frac{\sigma_1}{\epsilon_0},
\label{BC1}
\end{equation}
and 
\begin{equation}
\left. \epsilon_p \frac{\partial \Phi_{\textrm{II}}(r)}{\partial r} \right |_{r = R+w} - 
\left. \epsilon_w \frac{\partial \Phi_{\textrm{III}}(r)}{\partial r} \right |_{r = R+w} = \frac{\sigma_2}{\epsilon_0 }.
\label{BC2}
\end{equation}
The four unknown coefficients, $A, B, C,$ and $D$ are obtained from the boundary conditions and have the form obtained previously in  \cite{Siber2007}
\begin{equation}
A=\mathcal{A}/ \Delta, \, \, \, B=\mathcal{B}/ \Delta, C=\mathcal{C}/ \Delta, \, \, \, D=\mathcal{D}/ \Delta, 
\end{equation}
where
\begin{align}
\Delta =& \epsilon_{0}\epsilon_{w}\{(\epsilon_{p} - \epsilon_{w})\kappa_D w^{2} + \epsilon_{p}\kappa_D R^{2} \nonumber\\
& +[\epsilon_{p}(1 + 2 \kappa_D R) - \epsilon_{w}(1 + \kappa_D R)] w + \kappa_D R \nonumber\\
& \times [\epsilon_{w} w^{2} \kappa_D + \epsilon_{p}R + \epsilon_{w}(1+\kappa_D R)w]\coth (\kappa_D R)\},
\end{align}
and
\begin{eqnarray}
\mathcal{A} &=& \epsilon_{w}\sigma_{1}R^{2}\kappa_D w^{2} + \epsilon_{p}[\sigma_{1}R^{2} + \sigma_{2}(R + w)^{2}]R,\nonumber\\
&& +\epsilon_{w}\sigma_{1}(1 + \kappa_D R)w R^{2}\text{csch} (\kappa_D R)\nonumber\\
\mathcal{B} &=& [\{[\epsilon_{p}\sigma_{1}R^{2} + \sigma_{2}(R + w)^{2}) - \epsilon_{w}\sigma_{2}(R + w)^{2}]w\}\nonumber\\
&& +\epsilon_{p}[\sigma_{1}R^{2} + \sigma_{2}(R + w)^{2}]R\nonumber\\
&& +\epsilon_{w}\sigma_{2}(R + w)^{2}\delta \kappa_D R \coth (\kappa_D R)]\exp [\kappa_D (R + w)],\nonumber\\
\mathcal{C} &=& \epsilon_{w}\{\sigma_{1}w^{2}\kappa_D R^{2} + [\sigma_{1}R^{2} + \sigma_{2}(R + w)^{2}]R\nonumber\\
&& +\sigma_{1}R^{2}(1 + 2\kappa_D R)w\nonumber\\
&& -\sigma_{2}(R + w)^{2}\kappa_D \coth (\kappa_D R)\},\nonumber\\
\mathcal{D} &=&\epsilon_{p}[\sigma_{1}R^{2} + \sigma_{2}(R + w)^{2}]\nonumber\\
&& -\epsilon_{w}[\sigma_{1}R^{2} + \sigma_{2}(R + w)^{2} + \sigma_{1}R^{2}\kappa_D (R + w)]\nonumber\\
&& +\epsilon_{w} \kappa_D \sigma_{2}(R + w)^{2}R \coth (\kappa_D R).
\label{consts}
\end{eqnarray}
The results quoted in the main text are based on the various limits stemming from  these expressions.

\subsection{Comparison between the full Poisson-Boltzmann and the approximate Debye-H\"uckel solutions \label{appDHPB}}

Here we compare the linearized DH solution with the full numerical solution of the PB equation. Technically this refers to solutions of the full PB equation
\begin{equation}
\frac{1}{r}\frac{d^2 (r\Phi(r))}{dr^2} + \frac{{\kappa_D}^2}{\beta e_0} \sinh{\beta e_0 \Phi(r)} = 0 \
\end{equation}
and the linearized DH equation
\begin{equation}
\frac{1}{r}\frac{d^2 (r\Phi(r)) }{dr^2} + {\kappa_D}^2 \Phi(r) = 0 
\end{equation}
in the regions (I and III) accessible to the electrolyte ions. The methodology for obtaining the PB numerics has been explained in details in our previous publications \cite{Maj18,Maj19,Maj20} and will not be elaborated here. 

In Figs. \ref{Fig8} and \ref{Fig9} we compare the spatial profile of $\textrm{pH}$ as obtained from the PB and the DH solutions. We notice that overall the difference for the chosen values of the parameters is small, but is smaller for $\alpha \leq 0$, Fig. \ref{Fig8}, than for $\alpha \geq 0$, Fig. \ref{Fig9}. In fact for large positive $\alpha$ the DH solution ceases to be a good approximation for the PB result, which would invalidate the DH approach. We also  specifically indicate the difference in the surface charge densities  $\Delta\sigma_i = \sigma_i^{\text{PB}} - \sigma_i^{\text{DH}}$ obtained from the two approaches in order to facilitate the comparison.

One conclusion following from the numerical results of the PB and DH approaches is that qualitatively they are very similar, but for certain values of the parameters there are quantitative differences. It seems that one can thus safely use the DH approximation if the focus is on the qualitative features, whereas a PB based calculation would be needed in order to do quantitative comparisons.

Concerning the quantitative mismatch between the PB and DH theories, we see the same trends as reported earlier \cite{Maj16}. For equal surface charge densities, linear DH theory overestimates the electrostatic potential. As our current study suggests, this remains true even when charge regulation is included unless the surface charge densities computed within the two theories do not vary too much. For significantly larger $\Delta\sigma_i$, as it is the case for the luminal region in Fig.~\ref{Fig8}(b), the electrostatic potential and the corresponding $\textrm{pH}$ can of course be larger for the PB theory.

\subsection{Curvature expansion parameters} \label{curvapp}

In writing down the DH electrostatic free energy in the curvature expanded form Eq.~\ref{fel} (see \cite{Sho16} for details), itself being based on the solution of the DH equation \cite{Siber2007}, we introduced the following quantities
\begin{widetext}
\begin{align}
f_0\left(\sigma_1, \sigma_2, \kappa_D, w \right) =\frac{\mu(\sigma_1+\sigma_2)^2+\kappa_D w(\sigma_1^2+\sigma_2^2)}{2\mu+\kappa_D w},
\end{align}
\begin{align}
f_1\left(\sigma_1, \sigma_2, \kappa_D, w \right)=&\kappa_D w\left(\frac{(3\mu+2(\kappa_D w)-1)\sigma_2^2+2\mu\sigma_1\sigma_2-(\mu-1)\sigma_1^2}{2\mu+(\kappa_D w)}\right),
\end{align}
and
\begin{align}
f_2\left(\sigma_1, \sigma_2, \kappa_D, w \right) =& \frac{\kappa_D w}{(2\mu+(\kappa_D w))^2} \bigg( (\mu-1)\left[(\kappa_D w)(\mu-1)-\mu\right]\sigma_1^2-2\mu((\kappa_D w)+1)(\mu-1)\sigma_1\sigma_2+ \nonumber\\
& + \left[(\kappa_D w)^3+(\kappa_D w)^2(4\mu-1)+(\kappa_D w)(5\mu^2-4\mu+1)-\mu(\mu-1)\right]\sigma_2^2\bigg)
\end{align}
\end{widetext}
that do not depend on the curvature of the membrane anymore.   In addition, for the curved membrane, $f_0, f_1$ and $f_2$ are not symmetric with respect to the two solvent accessible inner and outer surface charge densities.

\subsection{Dependence of $\Delta\textrm{pH}_{{\textrm{m}}}, \Delta\textrm{pH}_{{\textrm{in}}}$, $\Delta\textrm{pH}_{{\textrm{out}}}$ on curvature}

Taking into account the DH solution for $\Phi_{\textrm{I}}, \Phi_{\textrm{II}}$ and $\Phi_{\textrm{III}}$ above, and the definition of the  constants $A = A(R,w), B = B(R,w)$ and $C = C(R,w)$ in Eq. \ref{consts} that explicitly depend on $R, w$, we can rewrite the equations for the changes of the acidity Eqs.~\ref{DpHm}, \ref{DpHout} and \ref{DpHin} in the explicit form
\begin{align}
\Delta\textrm{pH}_{\textrm{m}} &= \textrm{pH}(R+w) -\textrm{pH}(R)\notag\\
&= \beta e_0\,\log_{10}{e}\left( B(R,w) \frac{\exp (-\kappa_{D} (R+w))}{(R+w)} \right.\notag\\
&\left.\phantom{=} - A(R,w) \frac{\sinh(\kappa_{D} R)}{R}\right),
\end{align}
\begin{align}
\Delta\textrm{pH}_{\textrm{out}} &= \textrm{pH}(r = R + w) - \textrm{pH}^0\notag\\
& = \beta e_0  \log_{10} e ~B(R,w) \frac{\exp (-\kappa_D (R+w))}{R+w},
\label{DeltapHout}
\end{align}
and
\begin{align}
\Delta\textrm{pH}_{\textrm{in}} &= \textrm{pH}(r = R) - \textrm{pH}^0\notag\\
& = \beta e_0 \log_{10} e ~A(R,w) \frac{\sinh(\kappa_D R)}{R}.
\end{align}

Figures \ref{Fig10} and \ref{Fig4gfw} show the plots of $\Delta\textrm{pH}_{\textrm{in}}$ vs $R$ with the rest of the parameters the same as before. Figure.~\ref{Fig4gfw} shows that  $\Delta\textrm{pH}_{\textrm{in}}$ decreases when we increase the size of the vesicle for $\alpha < 0$ (Fig.~\ref{Fig4gfw}(a)). While $\Delta\textrm{pH}_{\textrm{in}}$ increases when the size of the vesicle increases
for $\alpha > 0$ (Fig.~\ref{Fig4gfw}(b)). In addition, in Fig.~\ref{Fig10}, we found that the dependence of $\Delta\textrm{pH}_{\textrm{in}}$ on the radius, $R$, can be show increase, decrease or remain constant, depending on the values of the parameters $(\alpha,\chi)$. What is particularly interesting is the "critical isotherm"  corresponding to $\chi = -2\alpha$ case that shows no variation with radius up to a critical value and after that a non-monotonic dependence. {Figure. ~\ref{Fig10} shows the case of $\chi = 0$ (no surface interaction). The plots are for both $\alpha < 0$ (Fig.~\ref{Fig10}(b)) and $\alpha > 0$ (Fig.~\ref{Fig10}(c)). We found that the $\Delta\textrm{pH}_{\textrm{in}}$ are the same but different in sign.} 

\bibliographystyle{apsrev4-2}
\bibliography{pH_vesicle}

\end{document}